\documentclass[twocolumn,usenames,dvipsnames]{aastex62}
\usepackage{amsmath}
\usepackage{apjfonts}
\usepackage{amssymb}
\usepackage{xspace}
\usepackage{hyperref}
\usepackage{natbib}
\usepackage{multirow}
\usepackage{graphicx}
\usepackage{epstopdf}
\usepackage{epsf}
\usepackage{subfigure}
\usepackage{xcolor}
\usepackage{rotating}
\usepackage{mathrsfs}
\usepackage{slashbox}
\usepackage{multirow}
\usepackage{lipsum}
\usepackage[normalem]{ulem}

\def\sqrtJ{\sqrt{\Delta_r \Delta_z}}
\def\M1450{M_{\rm 1450}}
\def\CHIMP{~h^{-1}{\rm~Mpc}}

\def\lsim{\lesssim}
\def\Msun{{\rm M}_\odot}

\begin{document}

\title{Measuring the Density Fields around Bright Quasars at $z\sim 6$ with XQR-30 Spectra}

\author[0000-0002-3211-9642]{Huanqing Chen}
\affiliation{Department of Astronomy \& Astrophysics; 
The University of Chicago; 
Chicago, IL 60637, USA}

\author[0000-0003-2895-6218]{Anna-Christina Eilers}\thanks{NASA Hubble Fellow}
\affiliation{MIT Kavli Institute for Astrophysics and Space Research, 77 Massachusetts Ave., Cambridge, MA 02139, USA}

\author[0000-0001-8582-7012]{Sarah E.~I.~Bosman}
\affiliation{Max-Planck-Institut für Astronomie, Königstuhl 17, D-69117 Heidelberg, Germany}

\author[0000-0001-5925-4580]{Nickolay Y.\ Gnedin}
\affiliation{Theoretical Physics Department; 
Fermi National Accelerator Laboratory;
Batavia, IL 60510, USA}
\affiliation{Kavli Institute for Cosmological Physics;
The University of Chicago;
Chicago, IL 60637, USA}
\affiliation{Department of Astronomy \& Astrophysics; 
The University of Chicago; 
Chicago, IL 60637, USA}

\author[0000-0003-3310-0131]{Xiaohui Fan}
\affiliation{Steward Observatory, University of Arizona, 933 North Cherry Avenue, Tucson, AZ 85721, USA}

\author[0000-0002-7633-431X]{Feige Wang}
\thanks{NASA Hubble Fellow}
\affiliation{Steward Observatory, University of Arizona, 933 North Cherry Avenue, Tucson, AZ 85721, USA}

\author[0000-0001-5287-4242]{Jinyi Yang}
\altaffiliation{Strittmatter Fellow}
\affiliation{Steward Observatory, University of Arizona, 933 North Cherry Avenue, Tucson, AZ 85721, USA}

\author[0000-0003-3693-3091]{Valentina D'Odorico}
\affiliation{INAF-Osservatorio Astronomico di Trieste, Via Tiepolo 11, I-34143 Trieste, Italy}
\affiliation{Scuola Normale Superiore, Piazza dei Cavalieri 7, I-56126 Pisa, Italy}
\affiliation{IFPU-Institute for Fundamental Physics of the Universe, via Beirut 2, I-34151 Trieste, Italy}

\author[0000-0003-2344-263X]{George D. Becker}
\affiliation{Department of Physics \& Astronomy, University of California, Riverside, CA 92521, USA}

\author[0000-0002-4314-021X]{Manuela Bischetti}
\affiliation{INAF-Osservatorio Astronomico di Trieste, Via Tiepolo 11, I-34143 Trieste, Italy}

\author[0000-0002-5941-5214]{Chiara Mazzucchelli}\thanks{ESO Fellow}
\affiliation{European Southern Observatory, Alonso de Cordova 3107, Vitacura, Region Metropolitana, Chile}

\author[0000-0003-3374-1772]{Andrei Mesinger}
\affiliation{Scuola Normale Superiore, Piazza dei Cavalieri 7, 56126 Pisa, Italy}

\author[0000-0002-7129-5761]{Andrea Pallottini}
\affiliation{Scuola Normale Superiore, Piazza dei Cavalieri 7, 56126 Pisa, Italy}


\correspondingauthor{Huanqing Chen}
\email{hqchen@uchicago.edu}

\begin{abstract}
Measuring the density of the intergalactic medium using quasar sightlines in the epoch of reionization is challenging due to the saturation of Lyman-$\alpha$  absorption. Near a luminous quasar, however, the enhanced radiation creates a proximity zone observable in the quasar spectra where the Lyman-$\alpha$ absorption is not saturated. In this study, we use $10$ high-resolution ($R\gtrsim 10,000$) $z\sim 6$ quasar spectra from the extended XQR-30 sample to measure the density field in the quasar proximity zones. We find a variety of environments within $3$ pMpc distance from the quasars. We compare the observed density cumulative distribution function (CDF) with models from the \textit{Cosmic Reionization on Computers} simulation, and find a good agreement between $1.5$ to $3$~pMpc from the quasar. This region is far away from the quasar hosts and hence approaching the mean density of the universe, which allows us to use the CDF to set constraints on the cosmological parameter $\sigma_8=0.6\pm0.3$. The uncertainty is mainly due to the limited number of high-quality quasar sightlines currently available. Utilizing the more than $>200$ known quasars at $z\gtrsim 6$, this method will allow us in the future to tighten the constraint on $\sigma_8$ to the percent level. In the region closer to the quasar within $1.5$~pMpc, we find the density 
is higher than predicted in the simulation by $1.23 \pm 0.17$, suggesting the typical host dark matter halo mass of a bright quasar ($M_{\rm 1450}<-26.5$) at $z\sim 6$ is $\log_{\rm 10} (M/\Msun)=12.5^{+0.4}_{-0.7}$. 
\end{abstract}

\section{Introduction}
Discoveries of supermassive black holes within the first billion years of the universe challenge our understanding of the formation of the first quasars. In order to reach the masses of $10^9 \Msun$ at $z\gtrsim7.5$ as measured in \citet{banados2018,yang2020,wang2021}, 
supermassive black holes (SMBH) at the centers of the first quasars must form or/and accrete via mechanisms that are not yet fully understood -- for example, the direct collapse into heavy seeds or super-Eddington accretion \citep[see][for a review]{inayoshi2020}.
To trigger these exotic mechanisms, stringent conditions on their environments are often required. For example, to form a massive SMBH seed from a supermassive star, the progenitorial molecular cloud needs to be kept warm to avoid fragmentation \citep{bromm2003,wise2008,regan2009,shang2010,choi2013,latif2013}. Some important channels of achieving this include an excess of Lyman-Werner bands radiation or fast galaxy assembly, both of which prefer a scenario where the quasars live in a dense ``proto-cluster'' environments \citep[e.g.,][]{lupi2021}. On the other hand, other theories suggest that in order to achieve high accretion all the way to the SMBH, mass overdensity may not be as important as other environmental factors like the tidal field \citep[e.g.][]{matteo2017}.

Therefore, to understand the formation mechanism of the first quasars, it is crucial to measure their large-scale environments. Recent studies have been tackling this problem by counting galaxies in the quasar fields \citep[see e.g.,][and the references therein]{garciavergara2021,simpson2014,habouzit2019,ota2018,banados13,kim2009,mcgreer2014}. 
However, because galaxies are biased tracers of the density field and may be impacted by the radiation feedback from the quasar \citep{kitayama2000,kashikawa2007,chen2020,bosman2020}, it is hard to estimate the real underlying density field using a few detected galaxies. Also, because the selection is done based on only a few wavebands, there could be  interlopers in the foreground, and the completeness of these sources is also hard to quantify. This calls for a new independent way to measure the quasar environments.

Recently, \citet{chen2021b} showed that using proximity zone spectra we can recover the density field along the quasar line of sight. This provides a new way to constrain the environment of first quasars in the line-of-sight direction.
In the recent decades, we have gained high resolution spectra, which show detailed features in the proximity zones \citep[see e.g.,][]{eilers2017}.
In this paper, we use a sample of high resolution, high signal-to-noise ratio (SNR) spectra from the XQR-30 (main + extended) sample to recover the density field surrounding $z\sim 6$ quasars. By comparing the observed density field with the \textit{Cosmic Reionization on Computers} (CROC) simulation \citep{gnedin2014,chen2021a}, we also put constraints on cosmological parameters and quasar properties at $z\sim 6$. 

This paper is organized as follows. In Section \ref{sec:method}, we describe the basics of density recovery method, the quasar sample and measurement process. In Section \ref{sec:results}, we show the recovered density field of the $10$ quasar sightlines in our sample and their CDFs. In Section \ref{sec:discussion}, we compare the observed CDF with our simulation, and investigate various factors that impact the CDF, and study how the CDF at different distance range helps us constrain the cosmological parameter $\sigma_8$, typical halo mass of quasar hosts and quasar lifetime. A summary is provided in Section \ref{sec:summary}.

\begin{table*}[]
\begin{tabular}{|c|c|c|c|c|c|c|c|}
\hline
Subsample & Name            & Redshift & Method & Redshift ref.          & $\M1450$ & $\M1450$ (NIR) ref.  & Res. [km/s]\\ 
(1)&(2)&(3)&(4)&(5)&(6)&(7)&(8)\\ \hline
main      & VDESJ0224-4711  & 6.5222   & {[}CII{]} & \citet{wang2021b}     & -26.67      & \citet{wang2021b}       &       26                           \\ \hline
main      & PSOJ217-16      & 6.1498   & {[}CII{]} & \citet{decarli2018}  & -26.93      & \citet{banados2016}          &                   26  \\ \hline
main      & PSOJ323+12      & 6.5872   & {[}CII{]} & \citet{venemans2020} & -27.08      & \citet{wang2021b}                           &      26   \\ \hline
main      & PSOJ359-06      & 6.1719   & {[}CII{]} & \citet{venemans2020} & -26.62      & \citet{schindler2020}              &       26         \\ \hline
extended  & CFHQSJ1509-1749 & 6.1225   & {[}CII{]} &  \citet{decarli2018}  & -26.56      & \citet{decarli2018}    &                        26     \\ \hline
extended  & PSOJ036+03      & 6.5405   & {[}CII{]} & \citet{venemans2020} & -27.26      & \citet{wang2021b}       &                    26      \\ \hline
extended  & SDSSJ0927+2001  & 5.7722   & CO        & \citet{wang2010}   & -26.76      & \citet{banados2016}                 &       22     \\ \hline
extended  & SDSSJ1306+0356  & 6.0330   & {[}CII{]} & \citet{venemans2020} & -26.70       & \citet{schindler2020}               &     22          \\ \hline
extended  & ULASJ1319+0950  & 6.1347   & {[}CII{]} & \citet{venemans2020} & -26.80       & \citet{schindler2020}   &                    22         \\ \hline
extended  & SDSSJ0100+2802  & 6.3269   & {[}CII{]} & \citet{venemans2020} & -29.02      & \citet{schindler2020}   &                   26        \\ \hline
\end{tabular}
\caption{Quasar sample of this study.}
\label{tab:qsoSample}
\end{table*}

\section{Method}\label{sec:method}

In this study, we use a sample of high SNR quasars at $z>6$ to measure the density field, following the method described in \citet{chen2021b}.
We briefly describe the method here and refer the readers to that paper for details.
The method is based on the robust assumption that in the proximity zone the intergalactic medium (IGM) is in ionization equilibrium:
\begin{equation}\label{eq:ionEquil}
\Gamma(d) n_{\rm HI}=\alpha n_{\rm HII} n_e=\alpha \left(\frac{n_e}{n_{\rm H}}\right) n^2_{\rm H},
\end{equation}
where $\Gamma$ is the ionization rate of H~{\small{I}} of the gas at the location with distance $d$ away from the quasar, $\alpha$ is the recombination coefficient of H~{\small{II}}, $n_{\rm HI}$, $n_{\rm HII}$, $n_{\rm H}$ and $n_e$ are  the number density of neutral hydrogen, ionized hydrogen, total hydrogen and free electrons, respectively. 
Inside the proximity zones, the quasar dominates the radiation field \citep{calverley2011,chen2021b}, and the IGM is completely ionized and transparent ($\Gamma(d)\propto d^{-2}$) for the majority of the sightlines. The recombination rate $\alpha$ can be treated as independent of density to the first order in the very aftermath of reionization. 
The number density of neutral hydrogen is proportional to the optical depth at the corresponding pixel. Thus we have

\begin{equation}\label{eq:tau}
    {\tau} = {\rm const \times}\frac{\alpha }{ \Gamma(d)}   \left(\frac{n_e}{n_{\rm H}}\right) n^2_{\rm H}.
\end{equation}

To obtain the density field, we only need to factor out the ``constants'' $\Gamma (d)$, $\alpha$, and $n_e/n_{\rm H}$ by dividing this equation to the same equation for a baseline model.

Such a baseline model is obtained by running a 1D radiative transfer (RT) code on a uniform sightline. The 1D RT code is described in \citet{chen2021a}, which uses adaptive timestep to evolve the ionization state of hydrogen and helium, as well as gas temperature.
For each quasar, we create a sightline of the uniform IGM density at the cosmic mean at the corresponding redshift. The initial values of the IGM we adopt before the quasar turns on are, based on the mean value of the CROC simulation, $x_{\rm HI}=10^{-4}$, $x_{\rm HeI}=10^{-4}$, $x_{\rm HeII}=0.9$, $T=10^4$ K. We then post-process the uniform sight line with the same quasar ionizing spectra of the observed quasar (Section \ref{sec:ionflux}). We set the quasar lifetime to be $30$ Myr for the fiducial baseline model. This way we obtain the optical depth $\tilde{\tau}$ for this uniform sight line, and  the density field of the observed quasar is
\begin{equation}\label{eq:denRec}
\Delta \equiv 1+\delta=\sqrt{\frac{\tau}{\tilde{\tau}}},
\end{equation}
where ${\tau}$
is the observed optical depth for the  quasar and $\delta=\rho/\bar{\rho}-1$ is the cosmic overdensity. Note that the recovered density $\Delta$ is a geometric mean of real-space and redshift space density $\sqrtJ$ \citep{chen2021b}:
\begin{equation}\label{Eq:tildeDelta}
 \Delta=\sqrt{\Delta_r \Delta_z} \equiv  
\Delta_{r} \sqrt{\left| \frac{H}{H+dv_{\rm pec}/dr} \right|},
\end{equation}
where $\Delta_{r}$ is the real-space gas density (in units of the cosmic mean), $H$ is the Hubble constant at the quasar redshift, $v_{\rm pec}$ is the peculiar velocity along the line of sight, and $r$ is the proper distance from the quasar. As shown in \citet{chen2021b}, Equation \ref{eq:denRec} accurately recover the true density with a moderate scatter of $\lesssim 10\%$, due to some temperature fluctuation and slight deviations in radiation profiles from $r^{-2}$.

\subsection{Quasar Sample}

We use a sample of high spectral resolution and SNR quasars to measure the optical depth in the proximity zone and to recover the density field.
Currently, the most suitable sample for this task is the XQR-30 sample\footnote{http://xqr30.inaf.it/}, an ongoing survey of bright  $z>5.8$ quasars using the high-resolution X-Shooter spectrograph \citep{vernet2011} on the Very Large Telescope. The main sample of XQR-30 consists of 30 {quasars with new high-SNR observations}. The extended XQR-30 sample also includes 16 archival spectra taken with X-Shooter, which are reduced in a similar manner to the XQR-30 main sample, and have comparable or higher SNR. (\citealt{bosman2021}, {D'Odorico et al.~in prep}). {The SNRs of these spectra in the proximity zones, the regions we are interested in, are $\sim 100$.}

In order to avoid contamination of the density measurement by other absorption, we exclude quasars with broad absorption lines (BALs) or proximate  damped Lyman-$\alpha$ systems (p-DLA). Moreover, using Lyman-$\alpha$ spectra in the proximity zone to recover the density field requires an accurate measurement of the quasar redshift. 
Among all the redshift measurement techniques, sub-millimeter lines ([C~{\small{II}}] or CO) provide the most reliable and precise measurement. Because they are thought to trace the cold ISM, the systematic uncertainty is estimated to be smaller than $100$ km/s (equivalent to $\Delta z \approx 0.002$). With spectra measured from ALMA or NOEMA, the error (usually a few $10$ km/s) in fitting the sub-mm lines is far smaller than the systematic uncertainty. Therefore, we further constrain the sample by only including quasars with redshift measurements from  sub-millimeter [C~{\small{II}}] or CO lines. 
With these constraints, our final sample is listed in Table \ref{tab:qsoSample}. 

\begin{figure*}
    \centering
\includegraphics[width=0.99\textwidth]{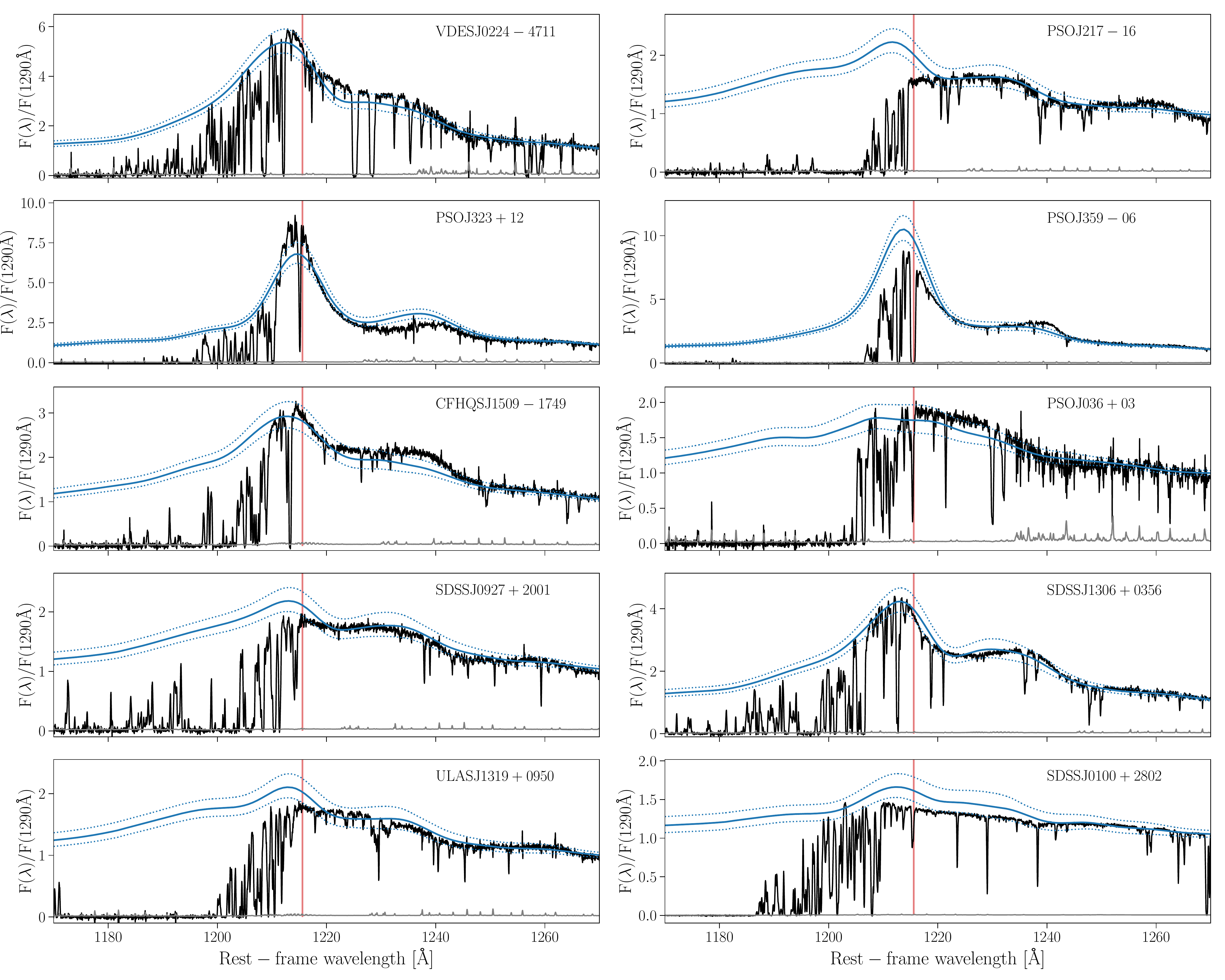}
    \caption{Rest-frame spectra for all the quasars in this study.  In each panel, the black line is the observed flux normalized to unity at $\lambda_{\text{rest}}=1290$\AA, the grey line is the noise. The solid blue line is the continuum fit, and the dotted blue lines show $1 \sigma$ uncertainty in continuum fitting. The vertical red line shows the position of $1215.6$ \AA , the wavelength of Lyman-$\alpha$. }
    \label{fig:continuum}
\end{figure*}

In the 6th column of Table \ref{tab:qsoSample}, we list the absolute magnitude $\M1450$ of each quasar at rest-frame wavelength $1450$ \AA, collected from literatures listed in the 7th column.
In the last column, we also list the average full width at half maximum (FWHM) spectral resolution determined from the measurement of the FWHM of the telluric lines in the X-Shooter single frames. The two reported values, 22 and 26 km/s corresponds to the different slit widths of 0.7 and 0.9 arcsec, respectively, adopted for the VIS arm (see D'Odorico et al.~in prep, for more details).    
At $z\sim 6$, the Hubble parameter is $H\approx 700$ km/s/pMpc. This means that with the high quality XQR-30 spectra, we can recover the density field  with the equivalent spatial resolution of $\sim 20$ pkpc.
The quasar spectra are shown in Figure \ref{fig:continuum}. {For details of the reduction procedure, see \citet{bosman2021,zhu2021} and {D'Odorico et al.~in prep}.}

\begin{figure}
    \centering
    \includegraphics[width=\columnwidth]{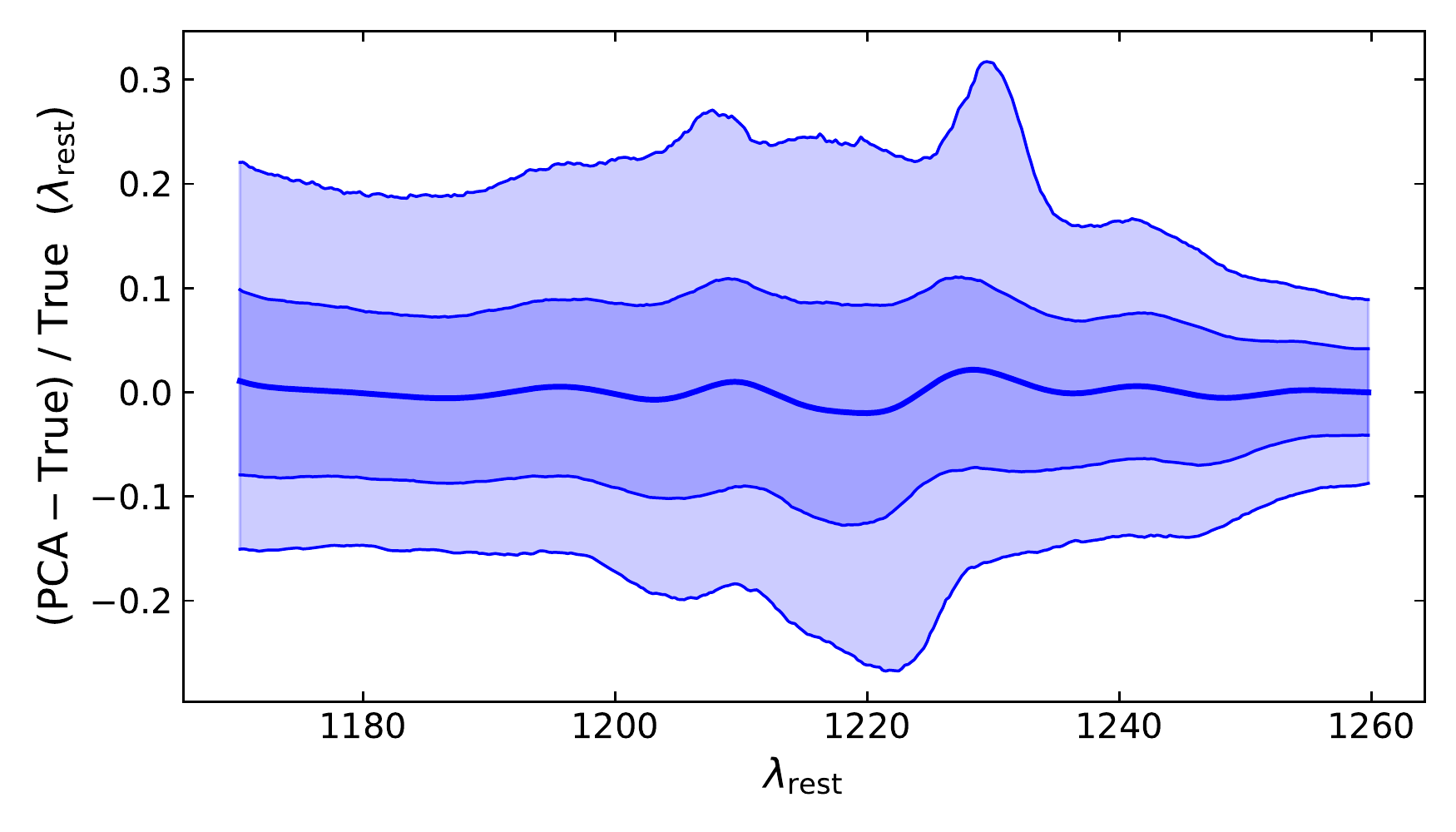}
    \caption{Performance of the PCA reconstruction on a testing sample of $7015$ eBOSS quasars at $2.25<z<3.5$. The shaded regions show the envelope containing $68\%$ and $95\%$ of deviations in the testing sample, at each wavelength. }
    \label{fig:PCA}
\end{figure}

\subsection{Continuum Fitting}\label{Sec: continuum}

We reconstruct the un-absorbed quasar emission in the Lyman-$\alpha$ line by performing a Principal Component Analysis (PCA) on a large set of quasars at lower $z$. Our procedure follows the log-PCA approach of \citet{davies2018} with the improvements introduced in \citet{bosman2021-pca}. 
{The specific PCA used in this paper differs from previous work mainly on the wavelength range we predict and the training/testing sample.} Here, we broadly summarise the procedure and highlight the ways in which it differs from previous work.

PCA reconstructions function by obtaining optimal linear decompositions of quasar spectra over two wavelength ranges: the unabsorbed `red side' of the spectrum (in this case $1220\,$\AA $<\lambda_{\rm rest}<2000\,$\AA) and the `blue side' which we wish to reconstruct (in this case $1170\,$\AA $<\lambda_{\rm rest}<1260\,$\AA). 
We use samples of low-$z$ quasars as training sets to learn the shape of intrinsic quasar Lyman-$\alpha$ emission. 
In practice, we use quasars selected from the SDSS-III Baryon Oscillation Spectroscopic Survey (BOSS) and the SDSS-IV Extended BOSS (eBOSS) DR14 \citep{BOSS,eBOSS,paris2018}. We select quasars with SNR$>10$ at $\lambda=1290\pm2.5$\AA, redshifts $2.25<z<3.5$ set by the visibility of the red and blue sides in the eBOSS spectral range, and which were not flagged as being BAL quasars. We added further quality checks to exclude quasars with missing data and identified further BALs by requiring that a smoothly-fit continuum do not drop below $0.7$ between $1290<\lambda<1570$\AA. Finally, we performed a visual inspection of the remaining quasars to exclude objects with strong proximate hydrogen absorption which precludes the recovery of the intrinsic Lyman-$\alpha$ emission line. The final sample consists of $14,029$ quasars. 

We fit both the red and blue sides of each quasar spectrum with a slow-varying spline function to which the PCA will be applied. This step enables the PCA to be less biased by random noise in the spectra. The spline is fit automatically following the procedure of \citet{young1979,carswell1982} as implemented by \citet{dallaglio2008} and refined by \citet{bosman2021-pca}. 
We then divide the sample randomly into a training and a testing sample of equal size.
We retain $15$ and $10$ components for the red-side and blue-side spectra, respectively. The projection between the two sides is obtained by dividing the weight matrices of the PCA components on both sides (e.g.~\citealt{paris2011}).

We evaluate the accuracy of the PCA prediction using the testing sample. The differences between PCA predictions and the automatically-fit splines at each rest-frame wavelength are encoded into a covariance matrix. Over the entire range of the blue side, the PCA prediction recovers the true continua within $+8.0/-8.2$\% ($+19.8/-16.4$\%) at $1\sigma$ ($2\sigma$). The reconstructions have a slight wavelength-dependent bias of $+0.7\%$ on average, which we correct for. Figure~\ref{fig:PCA} shows the mean bias and $\pm 1 \sigma, 2\sigma$ bounds of the PCA reconstructions performed on the testing sample. {Note that for each individual quasar, there may be an error of $\sim 10\%$ in continuum fit (Figure \ref{fig:continuum}), however, the bias of the whole sample should be $\approx 10\%/\sqrt{10}\approx 3\%$. We will study such systematics in Section \ref{sec:systematics}.}

\subsection{Ionizing spectra}\label{sec:ionflux}

To calculate the baseline $\tilde{\tau}$ necessary to recover the density field, we need to know the ionizing spectrum for each quasar. However, the ionizing part of quasar spectra is not directly observable, due to the large Gunn-Peterson optical depth at $z>4$. Therefore, we have to use the quasar templates developed at lower redshifts to convert the UV magnitude to ionizing flux.
We calculate the ionizing photon rate using Eq.~9 in \citet{runnoe2012},
$$ \log_{10} L_{\rm iso}=4.74+0.91 \log_{10}(1450 L_{1450}) , $$
to convert $M_{1450}$ to the isotropic bolometric luminosity. We then use the average breaking powerlaw ($L_\nu\propto \nu^{\alpha_\nu}$) spectral shape measured by \citet{lusso2015}, where
\begin{equation}
 \\
  \alpha_\nu =
    \begin{cases}
      -1.70 & \lambda<912 \text{\AA}\\
      -0.61 & \lambda>912 \text{\AA}\\
    \end{cases},
\end{equation}
to obtain the ionizing photon rate. For each quasar, we use this spectral shape with the corresponding ionizing photon rate to calculate the baseline model and obtain $\tilde{\tau}$. With this baseline $\tilde{\tau}$ and the observed $\tau$ described in the last subsection, the density field is obtained.

\begin{figure*}
    \centering
\includegraphics[width=0.99\textwidth]{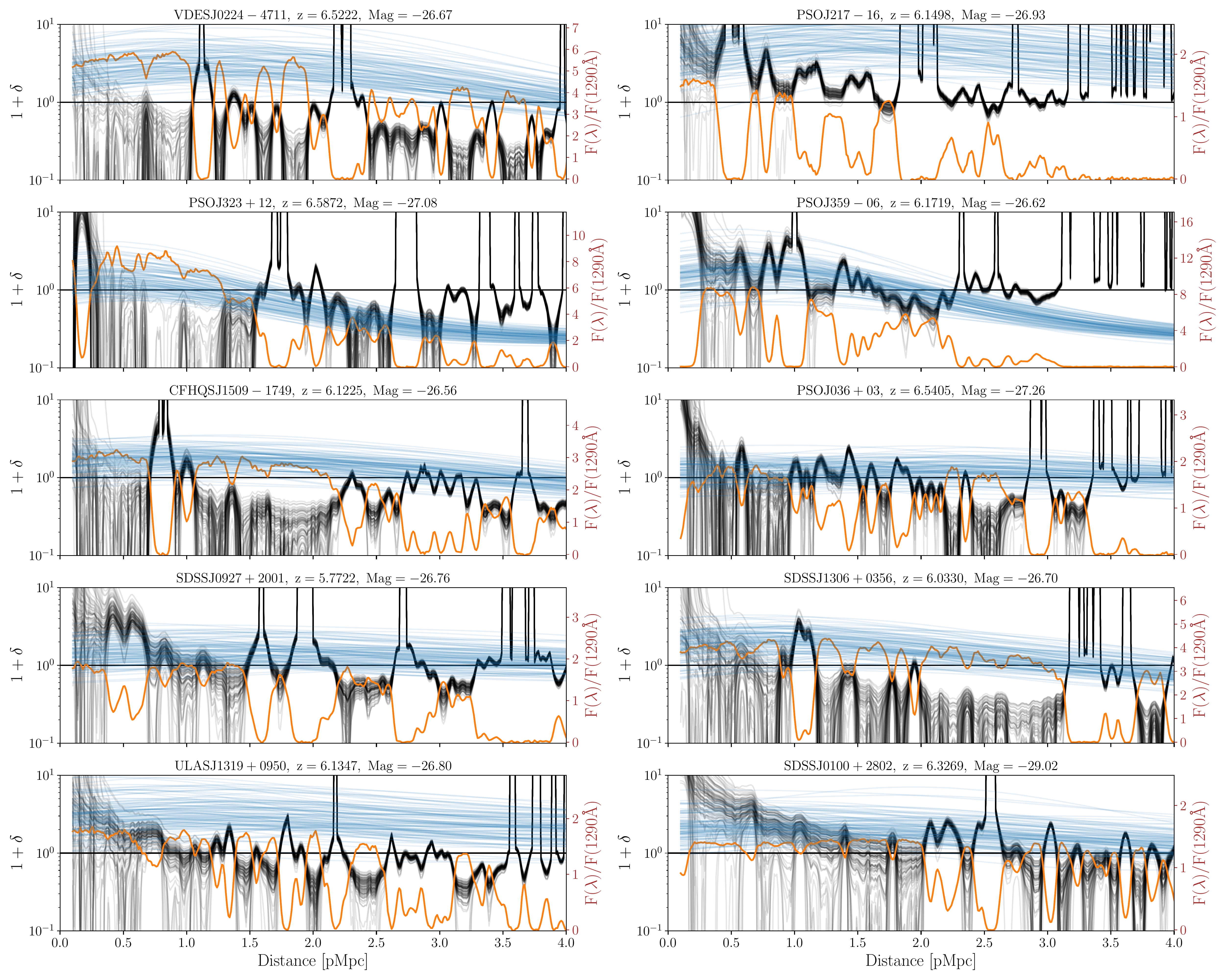}
    \caption{Each panel shows the density recovery result for each quasar. The orange line is the flux, while the thin blue lines are $100$ draws of  continuum fitting. The thin black lines are the corresponding recovered density field. 
    }
    \label{fig:conti_den}
\end{figure*}

\begin{figure*}
    \centering
\includegraphics[width=0.99\textwidth]{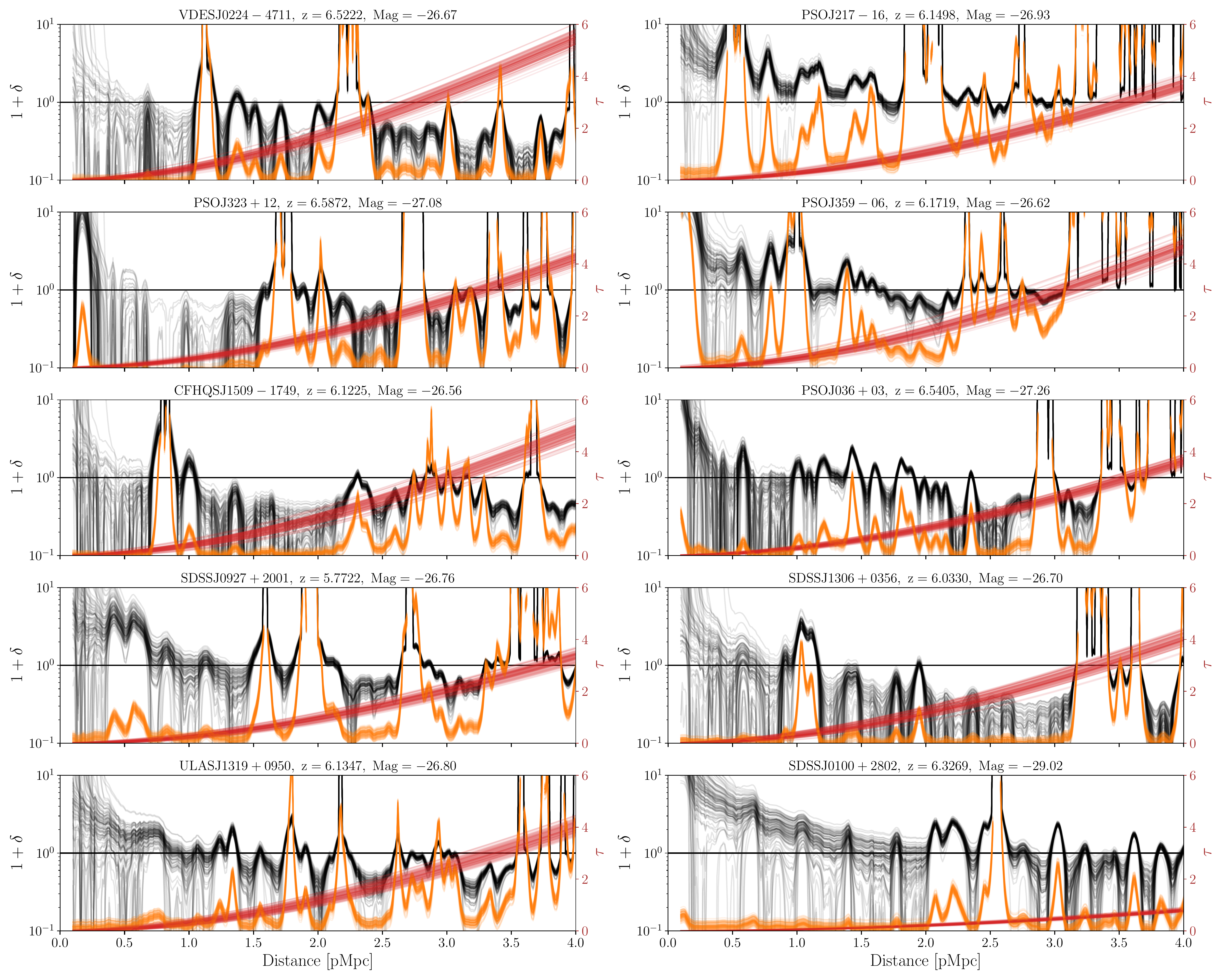}
    \caption{Similar to Figure \ref{fig:conti_den}, but here the thin orange lines are optical depths calculated from the  $100$ random draws of continuum fitting. The thin red lines show the optical depth model if the universe is uniform. The error is from the uncertainty of the exact quasar redshift. The thin black lines are the corresponding recovered density field, same as Figure \ref{fig:conti_den}.}
    \label{fig:model_den}
\end{figure*}

\section{Results}\label{sec:results}

\subsection{Recovered Density}
\label{sec:recDen}
In the density recovery process, we consider the error in $\tau$ from continuum fitting and the error in $\tilde{\tau}$ from quasar redshift measurement.  We use a correlation matrix to calculate the uncertainty in the continuum fitting (see Section \ref{Sec: continuum}), and assume that the error in redshift obeys a Gaussian distribution with $\sigma=\Delta z=0.002$. We use Monte-Carlo sampling to propagate these errors. For each quasar, we repeat the process 1000 times, every time drawing a random continuum and redshift to obtain $\tilde{\tau}$ and $\tau$ to compute density.

In Figure \ref{fig:conti_den} and Figure \ref{fig:model_den}  we show 100 out of 1000 realizations of the recovered density field of each quasar with black lines. The spikes are regions where Lyman-$\alpha$  absorption saturates. In Figure \ref{fig:conti_den} we show the observed flux in orange. The blue lines are 100 realizations of the continuum fittings. In Figure \ref{fig:model_den}, we show the 100 realizations of the observed optical depth ($\tau$) in orange, and the 100 realizations of modeled optical depth ($\tilde{\tau}$) in red. The uncertainty in observed optical depth (orange lines) is due to continuum fitting, while the uncertainty in modeled optical depth (red lines) is due to quasar redshift.

Certain regions have large uncertainties, especially underdense regions and regions very close to the quasar ($\lesssim 1$ pMpc). For the former the uncertainty is mostly due to continum fitting. The relative error in the observed optical depth is

$$ \frac{\Delta \tau}{\tau}=\frac{\ln (C+\Delta C)-\ln C}{\ln C - \ln F},$$
where $C$ is the continuum and $F$ is the observed flux. Since an underdense region corresponds to larger flux, a given error in quasar continuum results in a larger error in the recovered density in an underdense region as compared to an overdense region. Regions closer to the quasar also have relatively large uncertainties. This is because the baseline $\tilde{\tau}\propto d^2$, and the relative uncertainty due to the error in the quasar redshift $\Delta z$ is,

$$\frac{\Delta\tilde{\tau}}{\tilde{\tau}} \approx \frac{2 \Delta d_z}{d},$$
where $\Delta d_z\approx c\Delta z /(1+z) /H\approx 0.1$ pMpc. The relative uncertainty is thus larger at smaller distance.

One other thing to note concerns smoothing. Because of the instrumental broadening (FWHM$\approx 26$ km/s, equivalent to $40$ pkpc) of the sightlines, the recovered density field has an extra smoothing in addition to the intrinsic thermal broadening. Although we may not say too much about structures smaller than $40$ pkpc, however, larger-scale ($\gtrsim 200$ pkpc) structures should be robust. From Figure \ref{fig:conti_den} or Figure \ref{fig:model_den}, we can see a large variation in large-scale fields around quasars. It is typical to see an overdensity region of size $\lesssim 0.5$ pMpc, like the one at $2.0-2.5$ pMpc in VDESJ0224-4711. 
According to the CROC simulation \citep{chen2021b}, structures like this usually have $N_{\rm HI}\sim 10^{15}~\rm cm^{-2}$, neutral hydrogen fraction of $10^{-5}$, and most likely correspond to filaments of overdensity $\sim 10- 100$. There are also large voids of size $\sim 1$ pMpc, like the regions at $1-2$ pMpc in CFHQSJ1509-1749 and at $2-3$ pMpc in SDSSJ1306+0356.

\begin{figure*}
    \centering
    \includegraphics[width=0.45\textwidth]{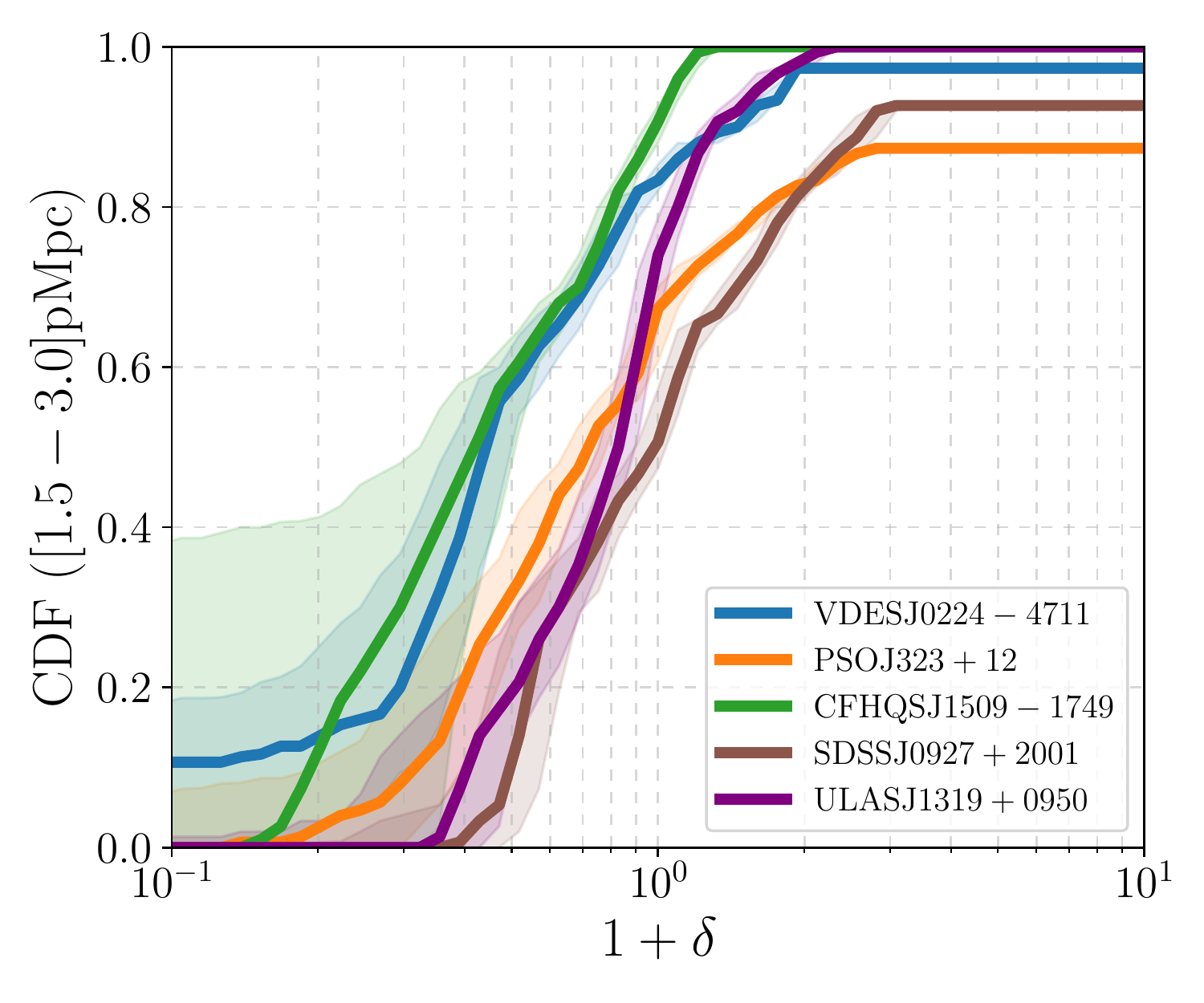}
    \includegraphics[width=0.45\textwidth]{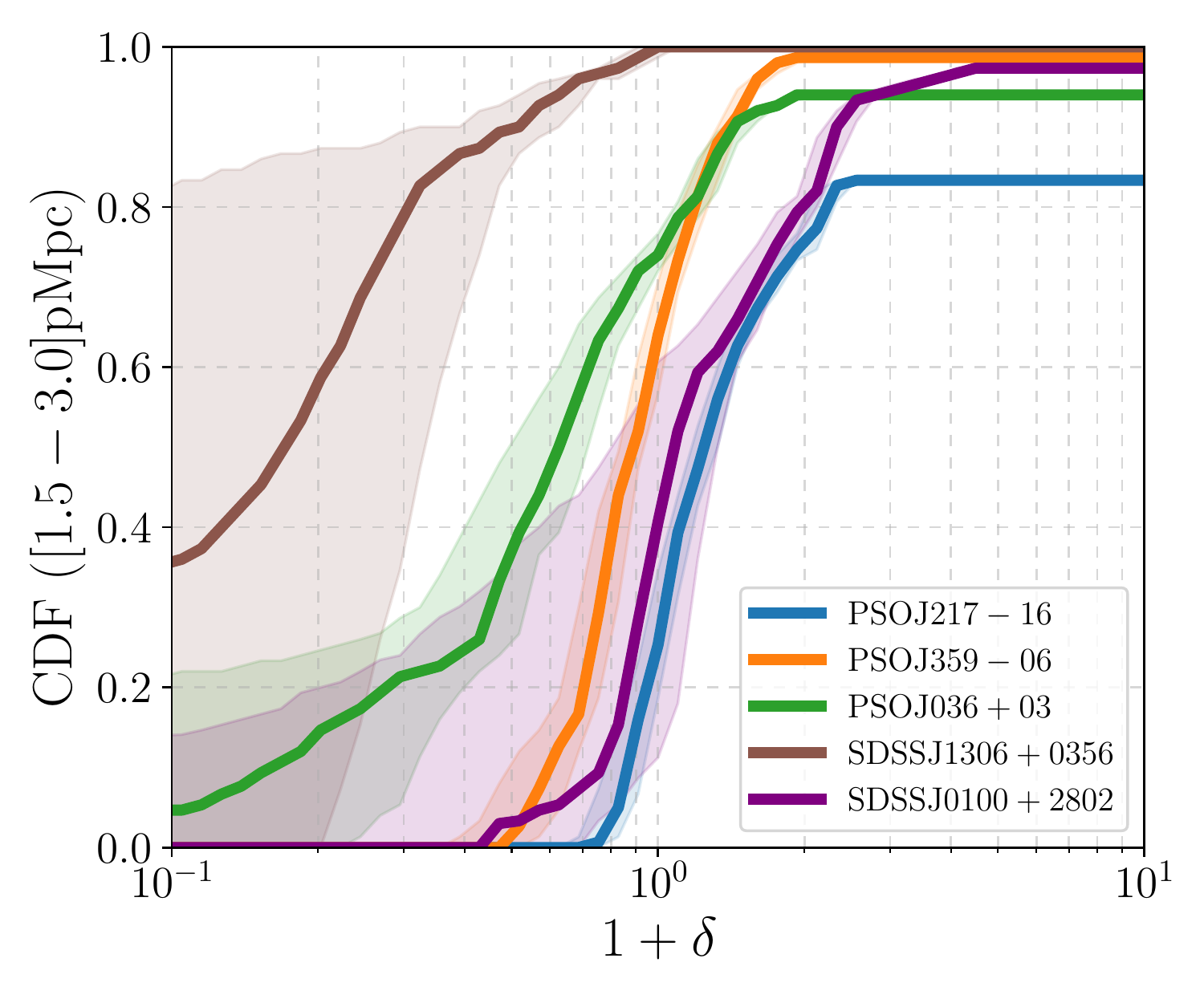}
    \caption{Cumulative distribution functions of density within $1.5-3.0$ pMpc for each quasar in the sample. Bands show $68\%$ uncertainty due to the uncertainties in the continuum fitting and the quasar redshift }
    \label{fig:individual_CDF}
\end{figure*}

\begin{figure}
    \centering
    \includegraphics[width=0.45\textwidth]{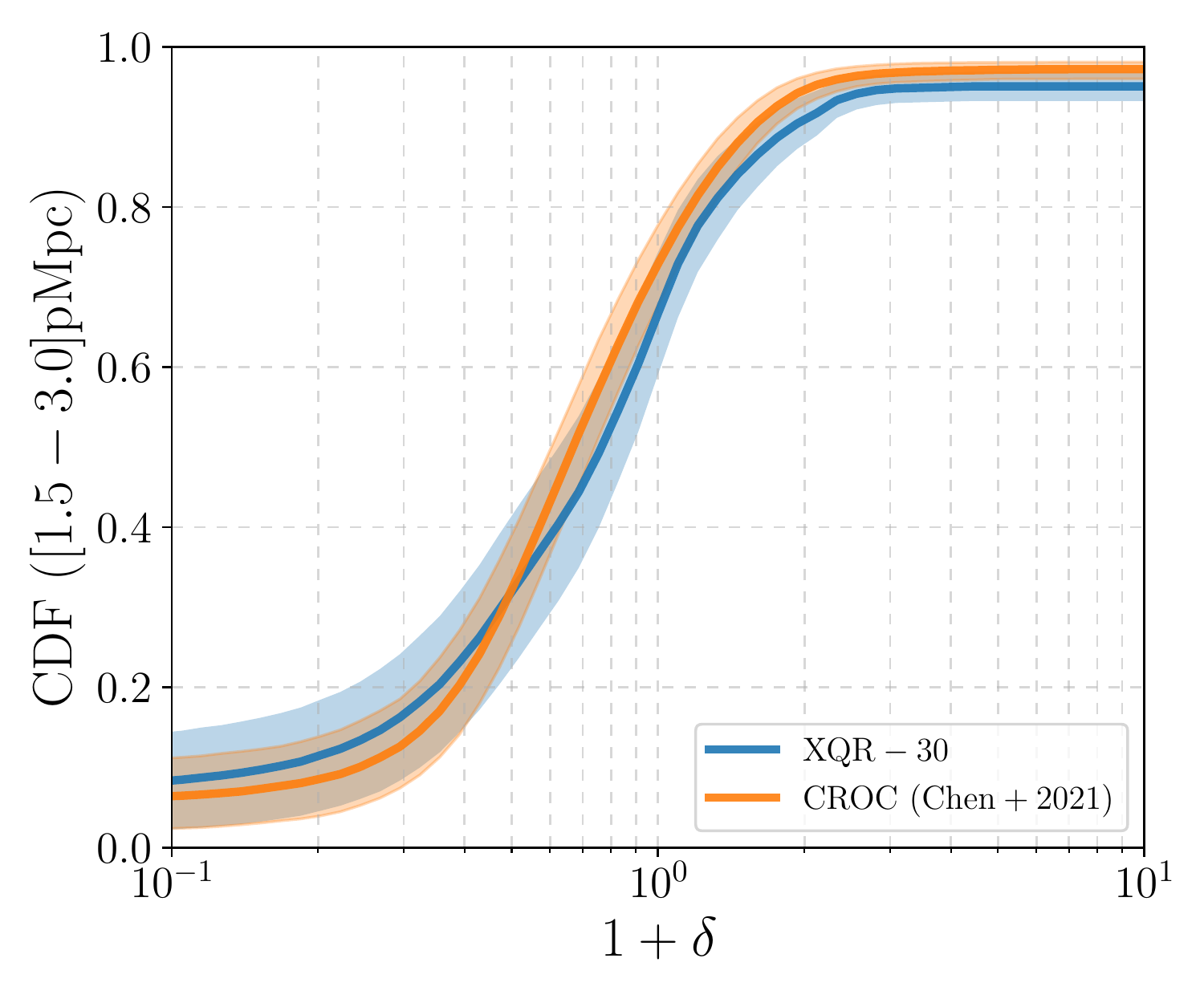}
\caption{Mean density CDF from the $10$ quasar sightlines in the XQR-30 sample (the blue line). The orange line is calculated from the synthetic spectra from the $40 \CHIMP$ box CROC simulation. The blue band shows $68\%$ uncertainty due to the errors in the continuum fitting and the quasar redshift, as well as the sample variance calculated by the jackknife estimator. The orange band shows $68\%$ uncertainty due to the uncertainty in the continuum fitting, the quasar redshift, the spectral noise, as well as the sample variance of 10 sightlines calculated from the total set of $6001$ sightlines from the $40 \CHIMP$ CROC simulation.
}
    \label{fig:compCROC}
\end{figure}

\begin{figure*}
    \centering
    \includegraphics[width=0.33\textwidth]{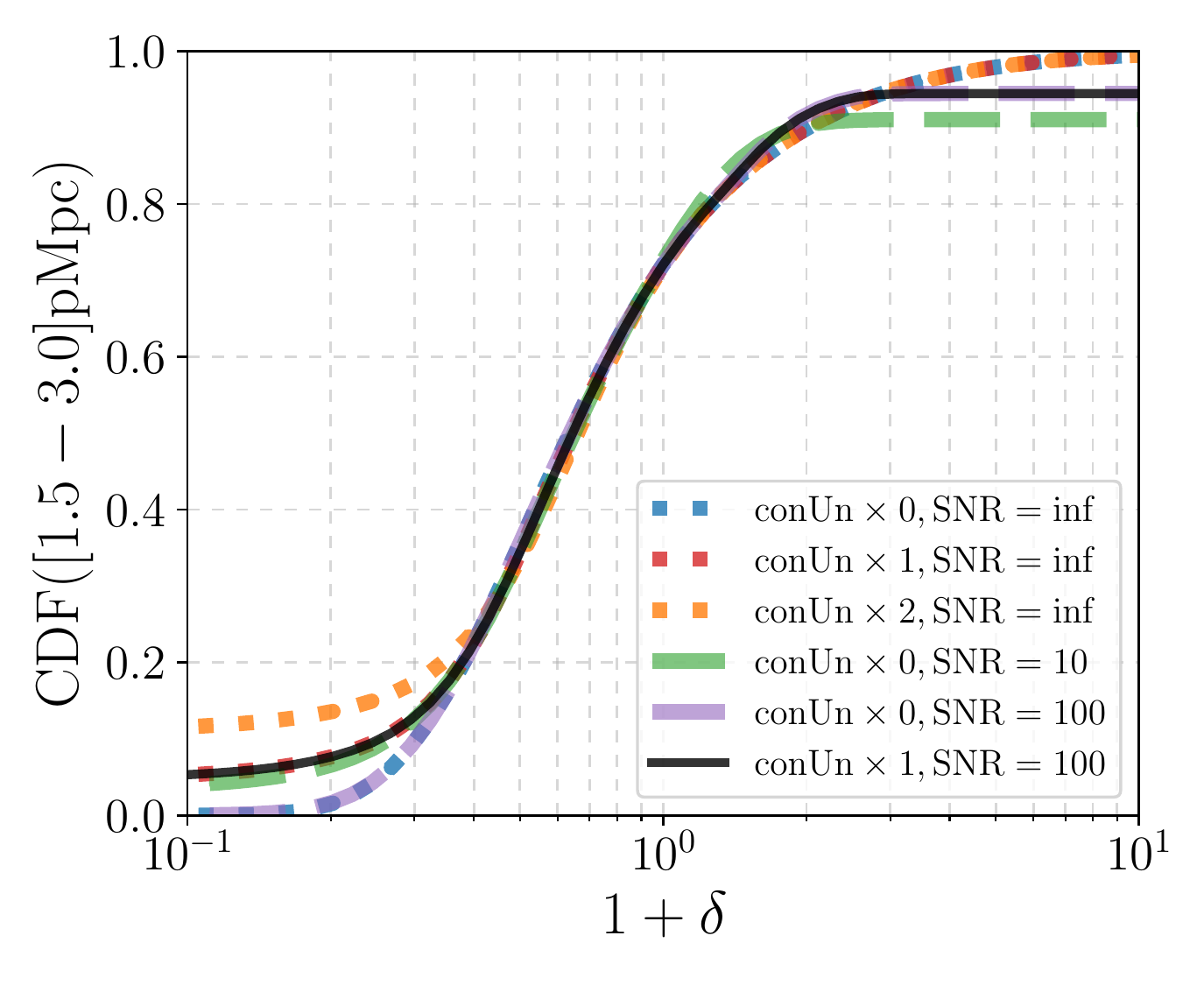}
    \includegraphics[width=0.33\textwidth]{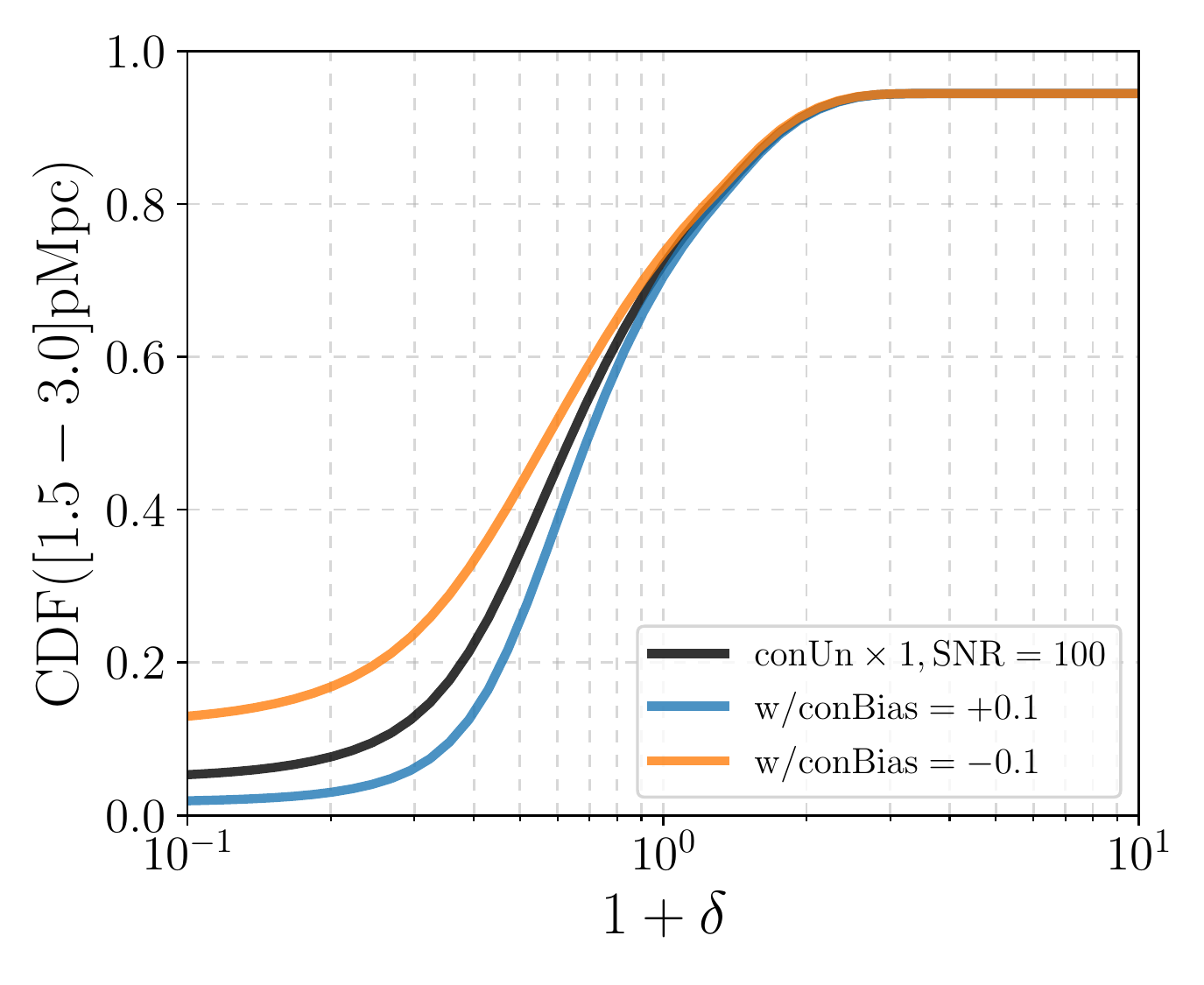}
    \includegraphics[width=0.33\textwidth]{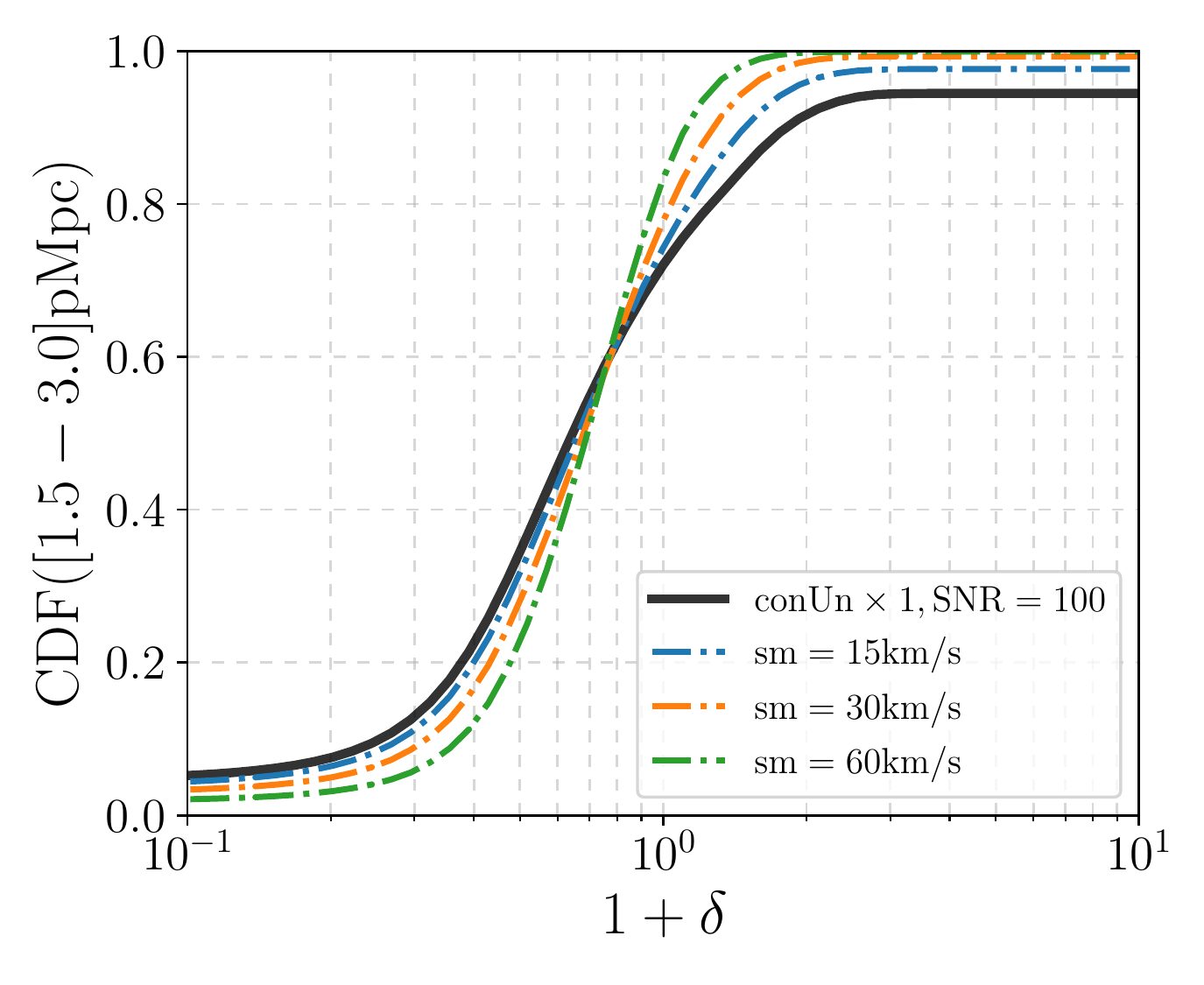}
    \caption{Left: effects of the continuum uncertainty and SNR on the CDF. The blue dotted line is assuming we know exactly the quasar continuum and the observation is perfect. The red and orange lines show how the CDF changes when the uncertainty in quasar continuum is included with $1\times$ or $2\times$ of our current estimate respectively. The solid green and purple lines are for the SNR of 10 or 100. The black line shows the CDF when we have both the continuum uncertainty and the finite SNR to match the actual XQR-30 data. The segment between 0.3-0.7 is robust to these uncertainties. Middle: the blue (orange) line shows the CDF when the continuum is biased high (low) by $10\%$. Right: spectra smoothing in observations steepen the CDF. The number listed for the dash-dotted lines show the equivalent Gaussian $\sigma$. }
    \label{fig:systematics}
\end{figure*}

\subsection{Cumulative Distribution Function (CDF) of Density}

The probability distribution function of the density is a simple and powerful tool to study the large-scale structure. In this section, we measure the density CDF for our XQR-30 sample. The CDF can help us constrain cosmological parameters and quasar properties, which we will discuss in the next section.

In Figure \ref{fig:individual_CDF}, we show the reconstructed density CDF in the distance range $1.5-3.0$ pMpc {(corresponding to $1211\,$\AA $ \gtrsim \lambda_{\rm rest}\gtrsim 1207$\AA)} for each quasar. We only choose such a narrow distance range, because closer to the quasar the density error is large due to the uncertainty on the quasar redshift, while farther from the quasar the ionization timescale is larger ($t_{\rm ion}\sim 0.1 $ Myr) and it is possible that the gas has not yet fully reached ionization equilibrium. Therefore, this is a very conservative choice to ensure the validity and accuracy of the density recovery procedure from Equation \ref{eq:denRec}. 

In Figure \ref{fig:individual_CDF}, we show the CDF for each quasar using different colors. Note that because instrumental broadening affects the shape of the CDF (see Section \ref{sec:systematics}), we apply additional Gaussian smoothings to bring the effective spectral resolution to FWHM$=30$ km/s for all quasars. Also note that when recovering the density with Equation \ref{eq:denRec}, some random draws of continuum may have flux lower than the observed flux at some wavelength, resulting in negative optical depth (Section \ref{Sec: continuum} and \ref{sec:recDen}). In such cases, we assign zero density to those pixels. Therefore, the CDFs in Figure \ref{fig:individual_CDF} do not always have CDF$\approx0$ at the very low density end ($1+\delta$=0.1). On the other hand, some pixels are saturated because of the spectral noise, and we assign an infinitely large value to such pixels. This is why there are plateaus at the high-density end in Figure \ref{fig:individual_CDF}.
From Figure \ref{fig:individual_CDF}, we can see that at the cosmic mean density the majority of the CDFs vary between $0.4-0.8$. The only exception is SDSSJ$1306+0356$, which is significantly more underdense than others. This is mainly due to the voids at $2-3$ pMpc from this quasar (Figure \ref{fig:conti_den}). The faint bands show the $68\%$ uncertainty due to the uncertainty in the quasar continuum and quasar redshift. Again, at the low-density end, the uncertainty is large for all quasars because of the relatively large uncertainty in the continuum fitting. The variation from sightline to sightline is also large, as expected, because we have chosen a small region, resulting in large sample variance.

\section{Discussion}\label{sec:discussion}
The recovered density fields at $z\sim 6$ can help us understand both cosmology and quasar physics. In this section, we explore how to constrain cosmological parameters and quasar environment and their lifetime.

\subsection{Comparing the CDF with simulations}
Simulations help us understand the complex information encoded in the spectra. In this section, we compare the recovered density CDF from the observational sample to the simulated one from the CROC simulation.

It is reasonable to assume that quasar sightlines sample similar environments at $1.5 - 3$ pMpc ($\sim 10 - 20$ cMpc), a region far away from the halo where the halo-mass bias is small. 
In this case, using the mean of the CDFs from these $10$ quasars can greatly reduce the statistical uncertainty. In Figure \ref{fig:compCROC}, we show the mean CDF of the 10 quasar sightlines from our observational sample in blue. {To compute the uncertainty, we consider both the uncertainty from the measurement $\sigma_{\rm obs}$ and that from the sample variance $\sigma_{\rm jk}$. To compute $\sigma_{\rm obs}$, we calculate the mean CDF $1000$ times, each time using one realization for each of the $10$ quasars. We quote the $68\%$ range as $\sigma_{\rm obs}$. We use the jackknife estimator to calculate the sample variance. Specifically, we exclude one quasar at a time to calculate the CDF of $9$ quasars. Then we quote the $\sqrt{9} \times$ the standard deviation of these $10$ CDFs as the error from sample variance $\sigma_{\rm jk}$. This error also captures the error from the continuum bias of one quasar. In Figure \ref{fig:compCROC}, we present the total error $\sqrt{\sigma^2_{\rm obs}+\sigma^2_{\rm jk}}$ as the blue band. 
}

As a model for the mean CDF, we use the $6001$ sightlines drawn from one of the CROC simulations. {Here we briefly describe the simulation and refer interested readers to \citet{chen2021a} for more details. The CROC simulation we used in this study is run in a $40\CHIMP$ box using an adaptive mesh refinement code ART \citep{kravtsov1999, kravtsov2002, rudd2008}. The initial conditions are sampled on a uniform grid size with the cell size of $40 h^{-1} \rm ~ckpc$ and the peak spatial resolution during the simulation is maintained at $100$ pc in proper units. The simulation includes physical processes like radiation-field-dependent gas heating and cooling, star formation and stellar feedback. The stars are the main driving source of reionization, and the radiation transfer is fully coupled with gas dynamics using OTVET algorithm \citep{gnedin01}. 
There are no individual quasars as ionizing sources in the simulation box, although the background radiation from the population of quasars is included. To model the quasar proximity zone spectra, we randomly draw sightlines centered on the $63$ most massive halos with total mass (dark matter + baryon) $M_h > 1.8\times 10^{11} \Msun$ at $z=6.1$ \citep{chen2021a}, and then run a time-dependent 1D RT code with quasar spectra.}

To conduct an apples-to-apples comparison, we re-run all sightlines with the properties of each observed quasars in our sample, folding in instrumental factors like spectral resolution and noise. Specifically, we post-process the sightlines with a spectral index of $\alpha_\nu = -1.7$ and a grid of quasar ionizing luminosities $\dot{N}_{\rm tot}=1\times10^{57}, 1.5\times10^{57}, 2\times10^{57},  2.5\times10^{57}, 3\times10^{57}, 3.5\times10^{57}, 4\times10^{57}, 1\times10^{58} \rm~ s^{-1}$.
We post-process all $6001 \times 8$ sightlines for {a fiducial quasar lifetime of} 30 Myr, and calculate the transmitted spectra.
For each observed quasar, we pick the luminosity in the grids that is closest to the observed luminosity.  We then multiply it with a random continuum drawn from the covariance matrix (Section \ref{Sec: continuum}), obtaining the modeled flux. We also add Gaussian noise to each pixel according to the average noise in the proximity zone of each quasar. We further use a Gaussian kernel of ${\rm FWHM}=30$ km/s to smooth the synthetic spectra so they have the same spectral resolution of those observed ones. We then recover the density using the exact same method as the observed one. Finally, we calculate the mean CDF of the $10$ sightlines and repeat the process $1000$ times to calculate the uncertainty.  We show this simulated CDF as the orange line in Figure \ref{fig:compCROC} and the $68\%$ uncertainty as the faint orange band. This uncertainty includes quasar continuum, quasar redshift, spectral noise, and sample variance from a $40~ \CHIMP$ box.

Comparing the blue and orange lines in Figure \ref{fig:compCROC}, we find that they agree very well. This agreement indicates that our assumptions (optically-thin proximity zones and ionization equilibrium of the IGM) that go into the density reconstruction are correct.

\subsection{Factors that impact the observed CDF}\label{sec:systematics}

There are several factors that impact the accuracy of the recovered density, including uncertainties in the continuum fitting, observational noise, limited spectral resolution, and uncertainty in the quasar ionizing flux. These uncertainties affect different parts of the CDF.

Among these, we have a relatively good handle on the continuum fitting and the noise level.
In the left panel of Figure \ref{fig:systematics}, we show how the scatter in the continuum fitting and the SNR affect the CDF. All the CDFs shown in this section are the mean using the $6001$ sightlines drawn from the same simulation and run with a typical ionizing photon rate $N_{\rm tot}=1.5\times 10^{57} ~\rm s^{-1}$. The blue dotted line is for a perfect instrument, i.e.\ with infinite spectral resolution and SNR, and for exactly known quasar continua. The red dotted line adds the realistic scatter (but no bias) in the continuum fitting (Section \ref{Sec: continuum}). Due to the scatter in the continuum fitting, in some pixels the continuum estimate will fall below the actual observed flux. Therefore, the optical depth becomes negative and the density cannot be recovered; we set the density to zero in such pixels with the result that the CDF does not approach zero in the limit of low $\delta$. If the uncertainty in the continuum fitting is twice larger, the value of the CDF at zero density increases further, as is shown by the orange dotted line. However, as long as there is no bias in the continuum fitting, with the continuum underestimate always balanced by an overestimate in some other pixels, the CDF values of $\gtrsim 0.3$ are not impacted. In the same panel, we also show how changing the SNR changes the CDF. We model the spectra with noise the same way as above, i.e.\ in each pixel we add and subtract Gaussian noise {at} $10\%$ of the continuum level randomly to model the SNR$=10$ (green dashed line). Limited SNR means there are saturated pixels where we cannot measure the true optical depth. As described earlier, for all saturated pixels we assign an infinitely large density and for pixels with a negative optical depth we assign  zero density. Adding {a} $10\%$ level of noise results in a plateau in the CDF at high densities due to the saturation, and also in an increase at the low density end. Increasing the SNR to $100$ (purple dashed line), {which is the SNR close to the actual XQR-30 sample}, results in a smaller plateau at the high density end and in no significant increase at the low density end. If both the continuum uncertainty and the SNR is at the XQR-30 level (but with infinite spectral resolution), the CDF is that traced by the black solid line, which has a non-zero value at the low density end and a plateau at the high density end. From this panel, we find that although the scatter in the continuum fitting and the spectral noise impact the low- and high- density ends of the CDF, the segment $0.3<$CDF$<0.7$ is robust.

While the uncertainty in the continuum fitting does not affect the CDF in the range $0.3-0.7$, the bias does. In the middle panel of Figure \ref{fig:systematics} we show how an average bias of $10\%$ in the continuum fitting affects the recovered CDF. Such a bias in the continuum impacts the lower density end more. If, for some reason, the continuum fitting method has an average bias of $10\%$ over the entire sample, then there will be a $\sim 10\%$ error in $1+\delta$ at CDF$=0.5$. We have tested our continuum reconstruction technique rigorously using lower redshift quasars, and found that the mean bias is $\approx 0.7\%$ (Figure \ref{fig:PCA}), well below $10 \%$ . However, {the bias level of this technique for very high-redshift quasars, especially for a relative small sample of them, is still to be investigated}. This potential bias is something to keep in mind when pushing the accuracy of the recovered CDF to the percent level.

In the right panel of Figure \ref{fig:systematics} we show how the spectral resolution affects the CDF. Thermal broadening has an effective Gaussian smoothing of $\sigma\approx 17$ km/s \citep{chen2021b}. If the spectral resolution is significantly smaller than this value, it becomes irrelevant.
Unfortunately, all real-life observations have limited spectral resolution, introducing some level of smoothing. 
Increasing the smoothing makes the CDF steeper and more "Heaviside-like". Also note, the smoothing from the finite spectral resolution is applied directly to the flux, which is an exponential function of optical depth. This non-linearity is the reason why the CDF recovered from the spectra with finite resolution does not have a mean density value at $\delta=0$.
Many saturated pixels can be smoothed out by a large smoothing kernel, so the saturation plateau at high densities also disappears.
Extra smoothing by the instrument downgrades the constraining power of the CDF. Therefore, to make improvements in constraining quasar properties with proximity zone spectra, it is crucial to have at least as high the spectral resolution as in the XQR-30 sample, and to accurately model the broadening effect from slit-width, seeing, spectrograph etc. {For X-Shooter, the uncertainty in spectral resolution is $\approx 10\%$. According to the right panel in Figure \ref{fig:systematics}, this level of uncertainty will not impact the conclusions in any significant way, as long as we use the part where CDF=$0.3-0.7$. }

The most uncertain quantity that we have to assume in the density recovery process is the ionizing radiation for each quasar. Because the ionizing part of the spectrum is heavily absorbed at $z\gtrsim 6$, we can not directly measure it. In this study we use the average quasar spectrum measured at $z\sim 2.4$ \citep{lusso2015}. We note that they quoted an uncertainty of $20\%$ in the mean ionization rate of H~{\small{I}}, while the scatter in the ionization rate of H~{\small{I}} between individual quasars may be a few times larger. Currently we have not found a good way to rigorously measure the uncertainty in this quantity. Perhaps machine learning techniques could in the future better predict the ionizing spectra using the red part of the quasar spectra, similar to how the continua are estimated. {Note that there is also an uncertainty in observed quasar magnitude, which is usually $\Delta {\rm mag}\approx 0.1$ in apparent magnitude. This translates to an uncertainty of $10\%$ in ionizing flux for a typical mag=$20$ quasar. Therefore, for a single quasar, the total uncertainty in ionizing radiation may be up to $50\%$. Because the optical depth depends on the square root of ionizing flux, this uncertainty translates up to $30\%$ on recovered density on a single quasar. Increasing the sample size can decrease such uncertainty, much as the same way as averaging out uncertainty in the continuum fitting. There are at least several hundred of bright quasars at $z\sim 6$, so the total uncertainty can go down to the percent level.}

\begin{figure}
    \centering
    \includegraphics[width=0.45\textwidth]{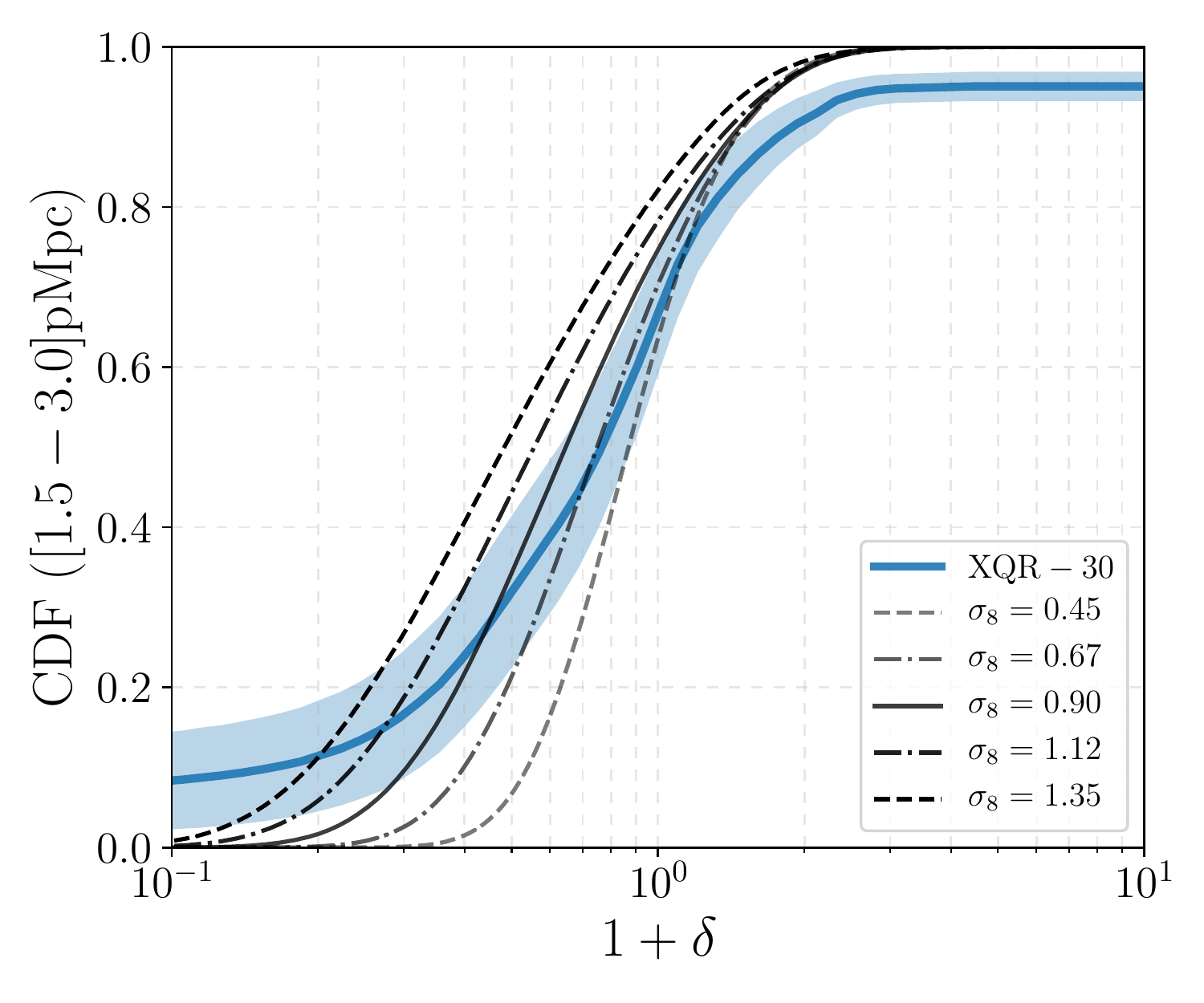}
\caption{Density CDF measured from the XQR-30 sample (blue line), overlapped by CDFs in cosmologies with different $\sigma_8$. The progressively darker shades represent larger $\sigma_8$.}
    \label{fig:sigma8}
\end{figure}

\begin{figure}
    \centering
    \includegraphics[width=0.45\textwidth]{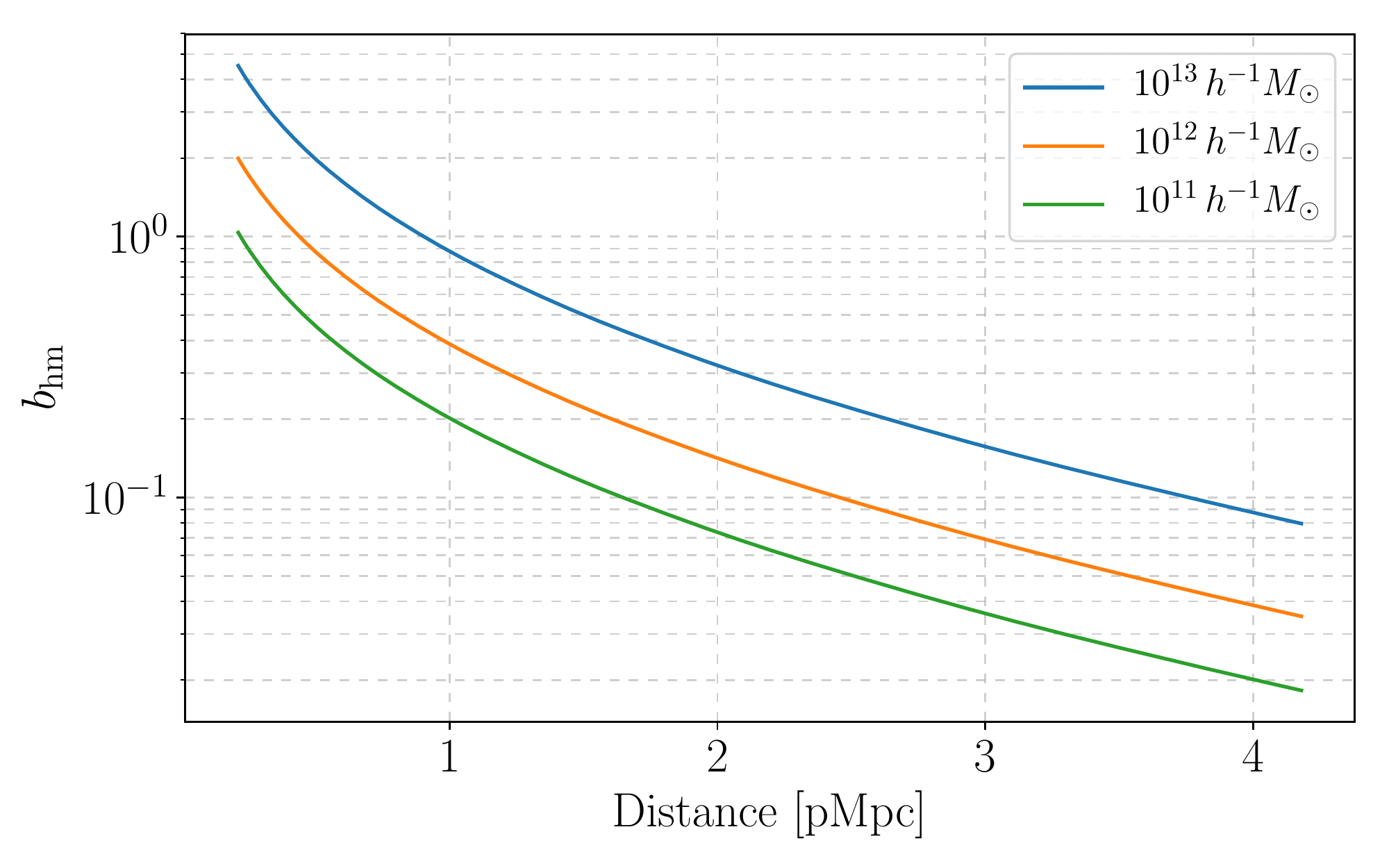}
    \caption{Halo-mass bias around halos of different masses at $z=6$ calculated using the python toolkit Colossus \citep{diemer2018}. }
    \label{fig:bias}
\end{figure}

\begin{figure}
    \centering
    \includegraphics[width=0.45\textwidth]{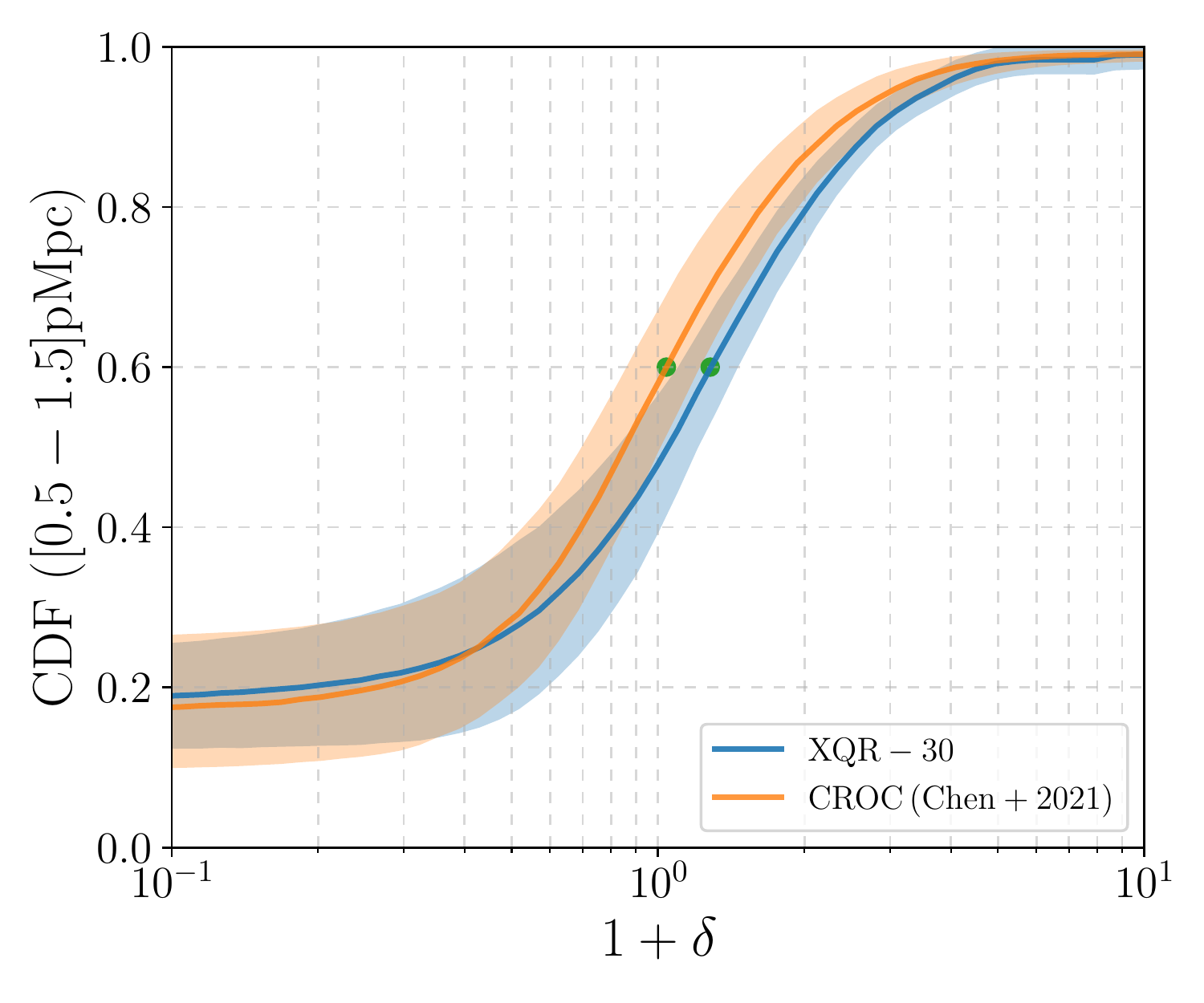}
    \caption{Same as Figure \ref{fig:compCROC}, but in an inner region of $0.5\sim 1.5$ pMpc, where the halo-mass bias is significant. The green dots mark locations where CDF=$0.6$, which corresponds to ($1+\delta$)=1.28 for XQR-30 and 1.04 for CROC, respectively.}
    \label{fig:compCROCinner}
\end{figure}

\begin{figure}
    \centering
    \includegraphics[width=0.45\textwidth]{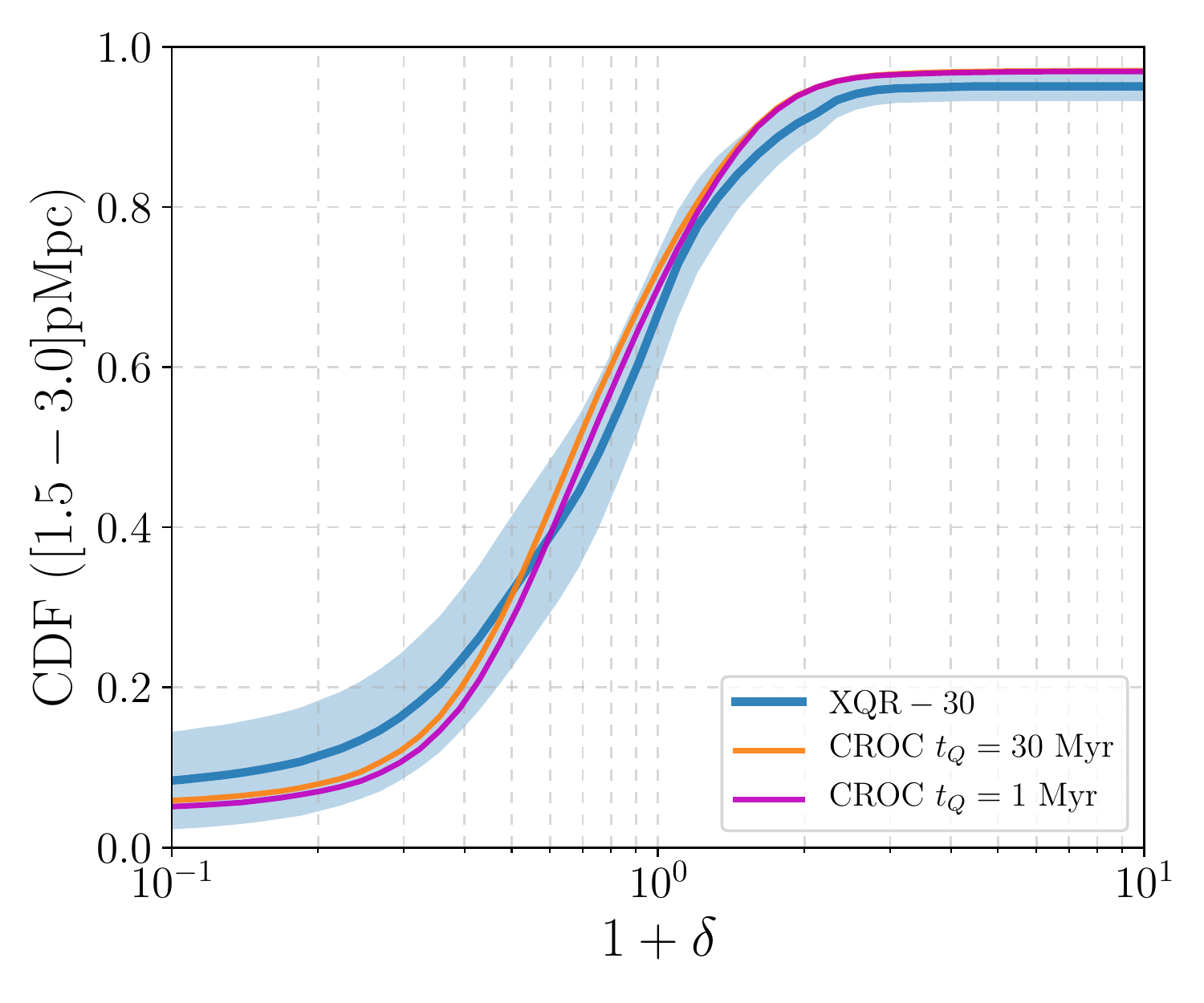}
    \caption{Density CDF in the distance range of $1.5 - 3.0$ pMpc. The blue line and orange lines are the same as in Figure \ref{fig:compCROC}. The magenta line is the recovered CDF from synthetic CROC spectra if the quasar lifetime is only $t_Q=1$ Myr.
    }
    \label{fig:compCROCtQ}
\end{figure}

\subsection{Application to Cosmology}
The measurement of the density field at $z\sim 6$ provides a unique way to study cosmology.
Currently, apart from the Cosmic Microwave Background (CMB), clustering measurements for matter, gas, or galaxies come from {the} low- or middle-redshift universe ($z\lsim 3)$, where the effect of the dark energy is not completely negligible. Different existing techniques have their own unique strengths and systematics. 
For example, measurements based on CMB lensing put strong constraints on a combination of $\Omega_m$ and $\sigma_8$, and, if combined with baryonic accoustic oscillation (BAO) measurements, they become  powerful for constraining the dark energy \citep[e.g.][]{wu2020}. At $z\sim 6$ cosmological measurements are extremely scarce. On the other hand, at this relatively high redshift the Universe is substantially different from the $z\lsim 3$ universe because it is completely matter dominated (at least in the absence of early dark energy). Therefore, studying the recovered density field at $z\sim 6$ can potentially help us understand cosmology better.

Cosmological parameters like  $\sigma_8$ (the amplitude of density fluctuations at $8h^{-1}$ cMpc scale) and $n_s$ (the spectral index of the matter power spectrum) control the shape of the density CDF \citep[e.g.,][]{chen2021c}. Therefore, our measurement can potentially be used to constrain these cosmological parameters.
In reality, however, the density PDF only mildly depends on $n_s$, and its sensitivity is mostly at the very low density end \citep{chen2021c}. Because this is also the place where measurement uncertainties are large, it is impossible to achieve meaningful constraints on $n_s$ with the current data. On the other hand, the density PDF as a whole is very sensitive to $\sigma_8$. The larger the $\sigma_8$, the wider the PDF, and the more the PDF peak shifts to the low density end. 

To constrain $\sigma_8$, we need to model the density PDF in cosmologies with different $\sigma_8$. We do so using the method described in \citet{chen2021c}. In that paper it is shown that one can approximate the PDF of the geometric mean of the real space and the redshift space density ($\sqrtJ$ - the quantity we recover from the proximity zones, see \S\ref{sec:method}) in $\Lambda$CDM cosmology with a set of self-similar simulations. The three parameters to describe the density PDF are the rms linear density fluctuation at a given smoothing scale, $\sigma$, the spectral index of the matter powers pectrum at that scale, $n_s$, and a parameter describing the redshift space distortions, $f$.
We consider the $\Lambda$CDM cosmology with different $\sigma_8$ at $z\sim6$. At this early time, the Universe is matter dominated, so $f=1$. The smoothing scale should be the equivalent thermal smoothing scale ($\sim 17$ km/s), which is $\approx 0.25$ comoving Mpc at $z\sim 6$. The spectral index at this scale is $n_s\approx -2$. And $\sigma$ is the parameter we vary to test the effect of $\sigma_8$. 
We thus calculate the density PDF from the self-similar simulations with $f=1$, $n_s=-2$ and a range of $\sigma=0.4 \sim 1.2$.

Because the recovered density field from observed quasars have extra smoothing due to limited spectral resolution, we
mimic this effect by doing the following. For each sightline drawn from the self-similar simulation, we calculate the transmitted flux $\tilde{F}=e^{-\tilde\tau\Delta_r\Delta_z}$. Since we do not have any way to compute the baseline model $\tilde\tau$ in self-similar simulations, we adopt $\tilde\tau=1$, which is a typical value for XQR-30 quasars in the distance range $1.5 - 3$ pMpc we focus on here. The result is a mock spectrum with infinite spectral resolution.
We then need to smooth it with a specific scale. Because the instrumental smoothing scale ($\sigma\approx 12~$km/s) is almost $0.75\times$ the intrinsic thermal broadening scale ($\sigma\approx 17~$km/s), we use a 1D Gaussian kernel of $0.75\times$ the size as the smoothing scale corresponding to the $\sigma$. Then from this smoothed transmitted flux $\tilde{F}$, we get back the synthetic recovered density by taking $\sqrt{-\ln{\tilde{F}}}$. This way we calculate the synthetic density CDFs for a grid of $\sigma=0.4,0.6,0.8,1.0,1.2$.
To translate these values of $\sigma$ to the cosmological parameter $\sigma_8$, one has two options. One is to analytically translate it between two spatial scales. However, it is complicated due to the involvement of both 1D and 3D smoothing. The other one, which we adopt, is to take a shortcut by calibrating them with the CROC simulation. The CROC simulation has $\sigma_8=0.8285$, and its CDF lies between $\sigma=0.6$ and $0.8$, agreeing the best with $\sigma=0.74$. Therefore, we multiply each $\sigma$ by a factor of $0.8285/0.74=1.12$ to obtain the effective $\sigma_8$. The five modeled CDFs with different $\sigma_8$ are shown as black lines in Figure \ref{fig:sigma8}.

In Figure \ref{fig:sigma8} we overplot the density CDFs of different $\sigma_8$ using thin black lines, with larger $\sigma$ represented by progressively darker shade. Note that when calculating the CDFs of different $\sigma_8$, we do not fully model the continuum fitting uncertainty and spectral noise. Therefore, we should only compare the section with $0.3<$CDF$<0.7$, which is robust against these factors (see the left panel of Figure \ref{fig:systematics}). Using the value at CDF$=0.6$, which is also robust against slight differences in smoothing scale, the constrain we obtain for $\sigma_8$ is $0.6\pm 0.3$, which is consistent with the concurrent cosmology measurement  \citep{planck2020,des2021}. {Compared to the Planck measurement of $\sigma_8=0.811 \pm 0.006$, our inferred value seems to lie in a lower end, which may be because the halo-bias is not completely negligible (see the next section and Figure \ref{fig:bias})} Also, as mentioned in the previous sections, there are other observational and physical factors that impact the observational CDF, like the quasar continuum fitting and the ionizing spectrum.
The large uncertainty mainly comes from the current limited sample size. {There are hundreds of bright quasars at $z\sim 6$, which means we can potentially reduce such uncertainty to $\approx 10\%$, rivaling current constraints from low-$z$ BAO measurements.} Developing an efficient procedure for jointly constraining them together with cosmological parameters requires substantially more research and is beyond the scope of this paper. That effort is, however, timely, as in the near future 30-meter class telescopes are expected to boost the number of available high-quality spectra by a factor of 30, paving the way to a likely breakthrough in our understanding of the quasar physics and cosmology from the proximity zone spectra.

\subsection{Constraining Properties of First Quasars}

The halo-mass bias ($b_{\rm hm}$) is very small at $\gtrsim 2$ pMpc (at $z\sim 6$) from a massive halo; however, it increases dramatically closer to the halo. In Figure \ref{fig:bias} we show $b_{\rm hm}$ as a function of distance around halos of different masses, calculated using the python toolkit Colossus\footnote{http://www.benediktdiemer.com/code/colossus/} \citep{diemer2018}. Around $1$ pMpc, for a halo of $10^{13} h^{-1} \Msun$, the bias is still around unity, but for halos of  $10^{12} h^{-1} \Msun$ or lower, the bias is less than $0.4$. Therefore, measuring the CDF at $\sim 1$ pMpc offers us a way to constrain the halo mass of the first quasars statistically.

In Figure \ref{fig:compCROCinner} we compare the CDF of the XQR-30 sample and the one from CROC in the inner region of $0.5 - 1.5$ pMpc. Compared with Figure \ref{fig:compCROC}, the uncertainty in the CDF of the XQR-30 sample in this distance range is slightly larger than that in the $1.5 - 3$ pMpc region, both because the distance range is smaller and the error in the quasar redshift has a larger impact at shorter distances. Also, compared with Figure \ref{fig:compCROC}, both curves shift to the right significantly. At CDF$=0.6$, in the outer region of $1.5 - 3.0$ pMpc, the XQR-30 sample (CROC sample) has the density $1+\delta\approx 0.9 (0.8)$, while in the inner region of $0.5 - 1.5$ pMpc the corresponding densities are $1+\delta=1.28$ for XQR-30 and $1+\delta=1.04$ for CROC, respectively. This bias agrees with the common assumption that bright quasars live in massive halos at $z\sim 6$. Furthermore, although this is not very statistically significant, there is a hint that the observed sample sits in a denser region on average than massive CROC halos --- at CDF$=0.6$ the density ratio is $1.23 \pm 0.17$. If we use this as the ratio of $(1+b_{\rm hm})_{\rm XQR-30}/(1+b_{\rm hm})_{\rm CROC}$, then considering that the halo masses for the halos we used in the CROC simulation are between $1.8\times 10^{11} ~\Msun$ and $1.8\times 10^{12} ~\Msun$ with the median of $2.3\times 10^{11} ~\Msun$, this indicates the typical mass of an XQR-30 quasar is $\log_{\rm 10} (M/\Msun)=12.5^{+0.4}_{-0.7}$.

Apart from constraining the halo mass, the recovered density can also potentially be used for constraining the quasar lifetime in the range of $1 - 100$ Myr.
In \citet{chen2021b}, it is shown that the quasar lifetime can bias the reconstructed density due to photoheating of He~{\small{II}}. For sufficiently short quasar lifetime $t_Q$, the ionized He~{\small{III}} bubble has not yet propagated beyond $3$ pMpc, resulting in cooler and thus slightly more neutral hydrogen gas. We model the recovered density CDF with the assumption that all of the $6001$ simulated CROC spectra have $t_Q=1$ Myr \citep[the average quasar lifetime at $z\sim6$ measured by][]{morey2021}, and show it with the magenta line in Figure \ref{fig:compCROCtQ}. There is a slight shift toward higher densities, and it is not statistically significant with just 10 quasar sightlines from the XQR-30 survey. However, in the future, after the 30-m class telescopes come online, we could potentially have more than hundred spectra with even higher spectral resolution. This improvement in data quantity and quality could help us to put constraints on the quasar lifetime in the Myr range. Also, the CDF may not be a particularly good statistics to constrain the quasar lifetime. Other statistical tools that we have not yet explored, such as the power spectrum, may be more useful in diagnosing a more subtle effect of spatial variation in thermal broadening, and this is a good topic for future work.

\section{Summary} \label{sec:summary}

Using a sample of high SNR, high-resolution $z\sim 6$ quasar spectra from the XQR-30 (main+extended) survey, we measure the density fields in their proximity zones out to $\sim 20$ cMpc for the first time. We compare the recovered density CDF with the modeled one from the CROC simulation in the outer region where the halo-mass bias is low, and find excellent agreement. We also test different factors that may impact the CDF, and find that the range between CDF$=0.3\sim 0.7$ is robust against uncertainties in the quasar continuum fitting and the spectral noise of the data.

We also explore how the recovered density PDF can constrain cosmology and quasar properties. We find that in the region $1.5- 3.0$ pMpc away from the quasars, using the sample of $10$ quasars one can reach a precision of $30\%$ in $1+\delta$ at CDF$=0.6$, although the uncertainty from quasar ionizing flux is an important systematic that requires further investigation. Using the CDF in that region, we can put a constraint on $\sigma_8=0.6\pm 0.3$. We also investigate the CDF in different distance ranges and find that the density field close to a quasar is systematically denser than in the outer region. The level of overdensity is comparable to the halo-mass bias of massive halos at $z\sim 6$. A comparison of the CDFs in the inner region between our observational sample and CROC simulation suggests the typical host halo mass of $M_{\rm 1450}<-26.5$ quasars at $z\sim 6$ is $\log_{\rm 10} (M/\Msun)=12.5^{+0.4}_{-0.7}$.

We also investigate the change in the recovered density CDF due to finite quasar lifetime. We find that the change is very small compared with the uncertainty of the existing XQR-30 data. However, we expect that the large increase in the quantity and quality of observational data from the future 30-meter class telescopes will lead to meaningful constraints on the quasar lifetime. In the future, we plan to investigate other statistical measures of the recovered density field, like the power spectrum, and to extract more information about reionization and first quasars.

\acknowledgements

H.C. thanks the support by NASA through the NASA FINESST grant NNH19ZDA005K.
ACE acknowledges support by NASA through the NASA Hubble Fellowship grant $\#$HF2-51434 awarded by the Space Telescope Science Institute, which is operated by the Association of Universities for Research in Astronomy, Inc., for NASA, under contract NAS5-26555.
XF and JY acknowleges suppot by the NSF through grant AST-1908284.
FW acknowledge support by NASA through the NASA Hubble Fellowship grant \#HF2-51448 awarded by the Space Telescope Science Institute, which is operated by the Association of Universities for Research in Astronomy, Incorporated, under NASA contract NAS5-26555.
GDB is supported by supported by the NSF through grant AST-1751404.
M.B. acknowledges support from PRIN MIUR project ``Black Hole winds and the Baryon Life Cycle of Galaxies: the stone-guest at the galaxy evolution supper'', contract \#2017PH3WAT.
AP acknowledges support from the ERC Advanced Grant INTERSTELLAR H2020/740120.
This manuscript has been co-authored by Fermi Research Alliance, LLC under Contract No. DE-AC02-07CH11359 with the U.S. Department of Energy, Office of Science, Office of High Energy Physics. Simulations used in this work were completed at the Argonne Leadership Computing Facility and the Blue Waters supercomputer. The Argonne Leadership Computing Facility is a DOE Office of Science User Facility supported under Contract DE-AC02-06CH11357. An award of computer time was provided by the Innovative and Novel Computational Impact on Theory and Experiment (INCITE) program. This research is also part of the Blue Waters sustained-petascale computing project, which is supported by the National Science Foundation (awards OCI-0725070 and ACI-1238993) and the state of Illinois. Blue Waters is a joint effort of the University of Illinois at Urbana-Champaign and its National Center for Supercomputing Applications.

\bibliographystyle{apj}
\bibliography{main}

\begin{thebibliography}{58}
\expandafter\ifx\csname natexlab\endcsname\relax\def\natexlab#1{#1}\fi

\bibitem[{{Ba{\~n}ados} {et~al.}(2013){Ba{\~n}ados}, {Venemans}, {Walter},
  {Kurk}, {Overzier}, \& {Ouchi}}]{banados13}
{Ba{\~n}ados}, E., {Venemans}, B., {Walter}, F., {Kurk}, J., {Overzier}, R., \&
  {Ouchi}, M. 2013, \apj, 773, 178

\bibitem[{{Ba{\~n}ados} {et~al.}(2016){Ba{\~n}ados}, {Venemans}, {Decarli},
  {Farina}, {Mazzucchelli}, {Walter}, {Fan}, {Stern}, {Schlafly}, {Chambers},
  {Rix}, {Jiang}, {McGreer}, {Simcoe}, {Wang}, {Yang}, {Morganson}, {De Rosa},
  {Greiner}, {Balokovi{\'c}}, {Burgett}, {Cooper}, {Draper}, {Flewelling},
  {Hodapp}, {Jun}, {Kaiser}, {Kudritzki}, {Magnier}, {Metcalfe}, {Miller},
  {Schindler}, {Tonry}, {Wainscoat}, {Waters}, \& {Yang}}]{banados2016}
{Ba{\~n}ados}, E., {Venemans}, B.~P., {Decarli}, R., {Farina}, E.~P.,
  {Mazzucchelli}, C., {Walter}, F., {Fan}, X., {Stern}, D., {Schlafly}, E.,
  {Chambers}, K.~C., {Rix}, H.~W., {Jiang}, L., {McGreer}, I., {Simcoe}, R.,
  {Wang}, F., {Yang}, J., {Morganson}, E., {De Rosa}, G., {Greiner}, J.,
  {Balokovi{\'c}}, M., {Burgett}, W.~S., {Cooper}, T., {Draper}, P.~W.,
  {Flewelling}, H., {Hodapp}, K.~W., {Jun}, H.~D., {Kaiser}, N., {Kudritzki},
  R.~P., {Magnier}, E.~A., {Metcalfe}, N., {Miller}, D., {Schindler}, J.~T.,
  {Tonry}, J.~L., {Wainscoat}, R.~J., {Waters}, C., \& {Yang}, Q. 2016, \apjs,
  227, 11

\bibitem[{{Ba{\~n}ados} {et~al.}(2018){Ba{\~n}ados}, {Venemans},
  {Mazzucchelli}, {Farina}, {Walter}, {Wang}, {Decarli}, {Stern}, {Fan},
  {Davies}, {Hennawi}, {Simcoe}, {Turner}, {Rix}, {Yang}, {Kelson}, {Rudie}, \&
  {Winters}}]{banados2018}
{Ba{\~n}ados}, E., {Venemans}, B.~P., {Mazzucchelli}, C., {Farina}, E.~P.,
  {Walter}, F., {Wang}, F., {Decarli}, R., {Stern}, D., {Fan}, X., {Davies},
  F.~B., {Hennawi}, J.~F., {Simcoe}, R.~A., {Turner}, M.~L., {Rix}, H.-W.,
  {Yang}, J., {Kelson}, D.~D., {Rudie}, G.~C., \& {Winters}, J.~M. 2018, \nat,
  553, 473

\bibitem[{{Bosman} {et~al.}(2021{\natexlab{a}}){Bosman}, {Davies}, {Becker},
  {Keating}, {Davies}, {Zhu}, {Eilers}, {D'Odorico}, {Bian}, {Bischetti},
  {Cristiani}, {Fan}, {Farina}, {Haehnelt}, {Kulkarni}, {Mesinger}, {Meyer},
  {Onoue}, {Pallottini}, {Qin}, {Ryan-Weber}, {Schindler}, {Walter}, {Wang}, \&
  {Yang}}]{bosman2021}
{Bosman}, S. E.~I., {Davies}, F.~B., {Becker}, G.~D., {Keating}, L.~C.,
  {Davies}, R.~L., {Zhu}, Y., {Eilers}, A.-C., {D'Odorico}, V., {Bian}, F.,
  {Bischetti}, M., {Cristiani}, S.~V., {Fan}, X., {Farina}, E.~P., {Haehnelt},
  M.~G., {Kulkarni}, G., {Mesinger}, A., {Meyer}, R.~A., {Onoue}, M.,
  {Pallottini}, A., {Qin}, Y., {Ryan-Weber}, E., {Schindler}, J.-T., {Walter},
  F., {Wang}, F., \& {Yang}, J. 2021{\natexlab{a}}, arXiv e-prints,
  arXiv:2108.03699

\bibitem[{{Bosman} {et~al.}(2020){Bosman}, {Kakiichi}, {Meyer}, {Gronke},
  {Laporte}, \& {Ellis}}]{bosman2020}
{Bosman}, S. E.~I., {Kakiichi}, K., {Meyer}, R.~A., {Gronke}, M., {Laporte},
  N., \& {Ellis}, R.~S. 2020, \apj, 896, 49

\bibitem[{{Bosman} {et~al.}(2021{\natexlab{b}}){Bosman},
  {{\v{D}}urov{\v{c}}{\'\i}kov{\'a}}, {Davies}, \& {Eilers}}]{bosman2021-pca}
{Bosman}, S. E.~I., {{\v{D}}urov{\v{c}}{\'\i}kov{\'a}}, D., {Davies}, F.~B., \&
  {Eilers}, A.-C. 2021{\natexlab{b}}, \mnras, 503, 2077

\bibitem[{{Bromm} \& {Loeb}(2003)}]{bromm2003}
{Bromm}, V. \& {Loeb}, A. 2003, \apj, 596, 34

\bibitem[{{Calverley} {et~al.}(2011){Calverley}, {Becker}, {Haehnelt}, \&
  {Bolton}}]{calverley2011}
{Calverley}, A.~P., {Becker}, G.~D., {Haehnelt}, M.~G., \& {Bolton}, J.~S.
  2011, \mnras, 412, 2543

\bibitem[{{Carswell} {et~al.}(1982){Carswell}, {Whelan}, {Smith}, {Boksenberg},
  \& {Tytler}}]{carswell1982}
{Carswell}, R.~F., {Whelan}, J.~A.~J., {Smith}, M.~G., {Boksenberg}, A., \&
  {Tytler}, D. 1982, \mnras, 198, 91

\bibitem[{{Chen}(2020)}]{chen2020}
{Chen}, H. 2020, \apj, 893, 165

\bibitem[{{Chen} \& {Gnedin}(2021{\natexlab{a}})}]{chen2021b}
{Chen}, H. \& {Gnedin}, N.~Y. 2021{\natexlab{a}}, \apj, 916, 118

\bibitem[{{Chen} \& {Gnedin}(2021{\natexlab{b}})}]{chen2021a}
---. 2021{\natexlab{b}}, \apj, 911, 60

\bibitem[{{Chen} {et~al.}(2021){Chen}, {Gnedin}, \& {Mansfield}}]{chen2021c}
{Chen}, H., {Gnedin}, N.~Y., \& {Mansfield}, P. 2021, arXiv e-prints,
  arXiv:2109.06194

\bibitem[{{Choi} {et~al.}(2013){Choi}, {Shlosman}, \& {Begelman}}]{choi2013}
{Choi}, J.-H., {Shlosman}, I., \& {Begelman}, M.~C. 2013, \apj, 774, 149

\bibitem[{{Dall'Aglio} {et~al.}(2008){Dall'Aglio}, {Wisotzki}, \&
  {Worseck}}]{dallaglio2008}
{Dall'Aglio}, A., {Wisotzki}, L., \& {Worseck}, G. 2008, \aap, 491, 465

\bibitem[{{Davies} {et~al.}(2018){Davies}, {Hennawi}, {Ba{\~n}ados}, {Simcoe},
  {Decarli}, {Fan}, {Farina}, {Mazzucchelli}, {Rix}, {Venemans}, {Walter},
  {Wang}, \& {Yang}}]{davies2018}
{Davies}, F.~B., {Hennawi}, J.~F., {Ba{\~n}ados}, E., {Simcoe}, R.~A.,
  {Decarli}, R., {Fan}, X., {Farina}, E.~P., {Mazzucchelli}, C., {Rix}, H.-W.,
  {Venemans}, B.~P., {Walter}, F., {Wang}, F., \& {Yang}, J. 2018, \apj, 864,
  143

\bibitem[{{Dawson} {et~al.}(2016){Dawson}, {Kneib}, {Percival}, {Alam},
  {Albareti}, {Anderson}, {Armengaud}, {Aubourg}, {Bailey}, {Bautista},
  {Berlind}, {Bershady}, {Beutler}, {Bizyaev}, {Blanton}, {Blomqvist},
  {Bolton}, {Bovy}, {Brandt}, {Brinkmann}, {Brownstein}, {Burtin}, {Busca},
  {Cai}, {Chuang}, {Clerc}, {Comparat}, {Cope}, {Croft}, {Cruz-Gonzalez}, {da
  Costa}, {Cousinou}, {Darling}, {de la Macorra}, {de la Torre}, {Delubac}, {du
  Mas des Bourboux}, {Dwelly}, {Ealet}, {Eisenstein}, {Eracleous}, {Escoffier},
  {Fan}, {Finoguenov}, {Font-Ribera}, {Frinchaboy}, {Gaulme}, {Georgakakis},
  {Green}, {Guo}, {Guy}, {Ho}, {Holder}, {Huehnerhoff}, {Hutchinson}, {Jing},
  {Jullo}, {Kamble}, {Kinemuchi}, {Kirkby}, {Kitaura}, {Klaene}, {Laher},
  {Lang}, {Laurent}, {Le Goff}, {Li}, {Liang}, {Lima}, {Lin}, {Lin}, {Lin},
  {Long}, {Lundgren}, {MacDonald}, {Geimba Maia}, {Malanushenko},
  {Malanushenko}, {Mariappan}, {McBride}, {McGreer}, {M{\'e}nard}, {Merloni},
  {Meza}, {Montero-Dorta}, {Muna}, {Myers}, {Nandra}, {Naugle}, {Newman},
  {Noterdaeme}, {Nugent}, {Ogando}, {Olmstead}, {Oravetz}, {Oravetz},
  {Padmanabhan}, {Palanque-Delabrouille}, {Pan}, {Parejko}, {P{\^a}ris},
  {Peacock}, {Petitjean}, {Pieri}, {Pisani}, {Prada}, {Prakash}, {Raichoor},
  {Reid}, {Rich}, {Ridl}, {Rodriguez-Torres}, {Carnero Rosell}, {Ross},
  {Rossi}, {Ruan}, {Salvato}, {Sayres}, {Schneider}, {Schlegel}, {Seljak},
  {Seo}, {Sesar}, {Shandera}, {Shu}, {Slosar}, {Sobreira}, {Streblyanska},
  {Suzuki}, {Taylor}, {Tao}, {Tinker}, {Tojeiro}, {Vargas-Maga{\~n}a}, {Wang},
  {Weaver}, {Weinberg}, {White}, {Wood-Vasey}, {Yeche}, {Zhai}, {Zhao}, {Zhao},
  {Zheng}, {Ben Zhu}, \& {Zou}}]{eBOSS}
{Dawson}, K.~S., {Kneib}, J.-P., {Percival}, W.~J., {Alam}, S., {Albareti},
  F.~D., {Anderson}, S.~F., {Armengaud}, E., {Aubourg}, {\'E}., {Bailey}, S.,
  {Bautista}, J.~E., {Berlind}, A.~A., {Bershady}, M.~A., {Beutler}, F.,
  {Bizyaev}, D., {Blanton}, M.~R., {Blomqvist}, M., {Bolton}, A.~S., {Bovy},
  J., {Brandt}, W.~N., {Brinkmann}, J., {Brownstein}, J.~R., {Burtin}, E.,
  {Busca}, N.~G., {Cai}, Z., {Chuang}, C.-H., {Clerc}, N., {Comparat}, J.,
  {Cope}, F., {Croft}, R. A.~C., {Cruz-Gonzalez}, I., {da Costa}, L.~N.,
  {Cousinou}, M.-C., {Darling}, J., {de la Macorra}, A., {de la Torre}, S.,
  {Delubac}, T., {du Mas des Bourboux}, H., {Dwelly}, T., {Ealet}, A.,
  {Eisenstein}, D.~J., {Eracleous}, M., {Escoffier}, S., {Fan}, X.,
  {Finoguenov}, A., {Font-Ribera}, A., {Frinchaboy}, P., {Gaulme}, P.,
  {Georgakakis}, A., {Green}, P., {Guo}, H., {Guy}, J., {Ho}, S., {Holder}, D.,
  {Huehnerhoff}, J., {Hutchinson}, T., {Jing}, Y., {Jullo}, E., {Kamble}, V.,
  {Kinemuchi}, K., {Kirkby}, D., {Kitaura}, F.-S., {Klaene}, M.~A., {Laher},
  R.~R., {Lang}, D., {Laurent}, P., {Le Goff}, J.-M., {Li}, C., {Liang}, Y.,
  {Lima}, M., {Lin}, Q., {Lin}, W., {Lin}, Y.-T., {Long}, D.~C., {Lundgren},
  B., {MacDonald}, N., {Geimba Maia}, M.~A., {Malanushenko}, E.,
  {Malanushenko}, V., {Mariappan}, V., {McBride}, C.~K., {McGreer}, I.~D.,
  {M{\'e}nard}, B., {Merloni}, A., {Meza}, A., {Montero-Dorta}, A.~D., {Muna},
  D., {Myers}, A.~D., {Nandra}, K., {Naugle}, T., {Newman}, J.~A.,
  {Noterdaeme}, P., {Nugent}, P., {Ogando}, R., {Olmstead}, M.~D., {Oravetz},
  A., {Oravetz}, D.~J., {Padmanabhan}, N., {Palanque-Delabrouille}, N., {Pan},
  K., {Parejko}, J.~K., {P{\^a}ris}, I., {Peacock}, J.~A., {Petitjean}, P.,
  {Pieri}, M.~M., {Pisani}, A., {Prada}, F., {Prakash}, A., {Raichoor}, A.,
  {Reid}, B., {Rich}, J., {Ridl}, J., {Rodriguez-Torres}, S., {Carnero Rosell},
  A., {Ross}, A.~J., {Rossi}, G., {Ruan}, J., {Salvato}, M., {Sayres}, C.,
  {Schneider}, D.~P., {Schlegel}, D.~J., {Seljak}, U., {Seo}, H.-J., {Sesar},
  B., {Shandera}, S., {Shu}, Y., {Slosar}, A., {Sobreira}, F., {Streblyanska},
  A., {Suzuki}, N., {Taylor}, D., {Tao}, C., {Tinker}, J.~L., {Tojeiro}, R.,
  {Vargas-Maga{\~n}a}, M., {Wang}, Y., {Weaver}, B.~A., {Weinberg}, D.~H.,
  {White}, M., {Wood-Vasey}, W.~M., {Yeche}, C., {Zhai}, Z., {Zhao}, C.,
  {Zhao}, G.-b., {Zheng}, Z., {Ben Zhu}, G., \& {Zou}, H. 2016, \aj, 151, 44

\bibitem[{{Dawson} {et~al.}(2013){Dawson}, {Schlegel}, {Ahn}, {Anderson},
  {Aubourg}, {Bailey}, {Barkhouser}, {Bautista}, {Beifiori}, {Berlind},
  {Bhardwaj}, {Bizyaev}, {Blake}, {Blanton}, {Blomqvist}, {Bolton}, {Borde},
  {Bovy}, {Brandt}, {Brewington}, {Brinkmann}, {Brown}, {Brownstein}, {Bundy},
  {Busca}, {Carithers}, {Carnero}, {Carr}, {Chen}, {Comparat}, {Connolly},
  {Cope}, {Croft}, {Cuesta}, {da Costa}, {Davenport}, {Delubac}, {de Putter},
  {Dhital}, {Ealet}, {Ebelke}, {Eisenstein}, {Escoffier}, {Fan}, {Filiz Ak},
  {Finley}, {Font-Ribera}, {G{\'e}nova-Santos}, {Gunn}, {Guo}, {Haggard},
  {Hall}, {Hamilton}, {Harris}, {Harris}, {Ho}, {Hogg}, {Holder}, {Honscheid},
  {Huehnerhoff}, {Jordan}, {Jordan}, {Kauffmann}, {Kazin}, {Kirkby}, {Klaene},
  {Kneib}, {Le Goff}, {Lee}, {Long}, {Loomis}, {Lundgren}, {Lupton}, {Maia},
  {Makler}, {Malanushenko}, {Malanushenko}, {Mandelbaum}, {Manera}, {Maraston},
  {Margala}, {Masters}, {McBride}, {McDonald}, {McGreer}, {McMahon}, {Mena},
  {Miralda-Escud{\'e}}, {Montero-Dorta}, {Montesano}, {Muna}, {Myers},
  {Naugle}, {Nichol}, {Noterdaeme}, {Nuza}, {Olmstead}, {Oravetz}, {Oravetz},
  {Owen}, {Padmanabhan}, {Palanque-Delabrouille}, {Pan}, {Parejko},
  {P{\^a}ris}, {Percival}, {P{\'e}rez-Fournon}, {P{\'e}rez-R{\`a}fols},
  {Petitjean}, {Pfaffenberger}, {Pforr}, {Pieri}, {Prada}, {Price-Whelan},
  {Raddick}, {Rebolo}, {Rich}, {Richards}, {Rockosi}, {Roe}, {Ross}, {Ross},
  {Rossi}, {Rubi{\~n}o-Martin}, {Samushia}, {S{\'a}nchez}, {Sayres}, {Schmidt},
  {Schneider}, {Sc{\'o}ccola}, {Seo}, {Shelden}, {Sheldon}, {Shen}, {Shu},
  {Slosar}, {Smee}, {Snedden}, {Stauffer}, {Steele}, {Strauss}, {Streblyanska},
  {Suzuki}, {Swanson}, {Tal}, {Tanaka}, {Thomas}, {Tinker}, {Tojeiro},
  {Tremonti}, {Vargas Maga{\~n}a}, {Verde}, {Viel}, {Wake}, {Watson}, {Weaver},
  {Weinberg}, {Weiner}, {West}, {White}, {Wood-Vasey}, {Yeche}, {Zehavi},
  {Zhao}, \& {Zheng}}]{BOSS}
{Dawson}, K.~S., {Schlegel}, D.~J., {Ahn}, C.~P., {Anderson}, S.~F., {Aubourg},
  {\'E}., {Bailey}, S., {Barkhouser}, R.~H., {Bautista}, J.~E., {Beifiori},
  A.~r., {Berlind}, A.~A., {Bhardwaj}, V., {Bizyaev}, D., {Blake}, C.~H.,
  {Blanton}, M.~R., {Blomqvist}, M., {Bolton}, A.~S., {Borde}, A., {Bovy}, J.,
  {Brandt}, W.~N., {Brewington}, H., {Brinkmann}, J., {Brown}, P.~J.,
  {Brownstein}, J.~R., {Bundy}, K., {Busca}, N.~G., {Carithers}, W., {Carnero},
  A.~R., {Carr}, M.~A., {Chen}, Y., {Comparat}, J., {Connolly}, N., {Cope}, F.,
  {Croft}, R. A.~C., {Cuesta}, A.~J., {da Costa}, L.~N., {Davenport}, J. R.~A.,
  {Delubac}, T., {de Putter}, R., {Dhital}, S., {Ealet}, A., {Ebelke}, G.~L.,
  {Eisenstein}, D.~J., {Escoffier}, S., {Fan}, X., {Filiz Ak}, N., {Finley},
  H., {Font-Ribera}, A., {G{\'e}nova-Santos}, R., {Gunn}, J.~E., {Guo}, H.,
  {Haggard}, D., {Hall}, P.~B., {Hamilton}, J.-C., {Harris}, B., {Harris},
  D.~W., {Ho}, S., {Hogg}, D.~W., {Holder}, D., {Honscheid}, K., {Huehnerhoff},
  J., {Jordan}, B., {Jordan}, W.~P., {Kauffmann}, G., {Kazin}, E.~A., {Kirkby},
  D., {Klaene}, M.~A., {Kneib}, J.-P., {Le Goff}, J.-M., {Lee}, K.-G., {Long},
  D.~C., {Loomis}, C.~P., {Lundgren}, B., {Lupton}, R.~H., {Maia}, M. A.~G.,
  {Makler}, M., {Malanushenko}, E., {Malanushenko}, V., {Mandelbaum}, R.,
  {Manera}, M., {Maraston}, C., {Margala}, D., {Masters}, K.~L., {McBride},
  C.~K., {McDonald}, P., {McGreer}, I.~D., {McMahon}, R.~G., {Mena}, O.,
  {Miralda-Escud{\'e}}, J., {Montero-Dorta}, A.~D., {Montesano}, F., {Muna},
  D., {Myers}, A.~D., {Naugle}, T., {Nichol}, R.~C., {Noterdaeme}, P., {Nuza},
  S.~E., {Olmstead}, M.~D., {Oravetz}, A., {Oravetz}, D.~J., {Owen}, R.,
  {Padmanabhan}, N., {Palanque-Delabrouille}, N., {Pan}, K., {Parejko}, J.~K.,
  {P{\^a}ris}, I., {Percival}, W.~J., {P{\'e}rez-Fournon}, I.,
  {P{\'e}rez-R{\`a}fols}, I., {Petitjean}, P., {Pfaffenberger}, R., {Pforr},
  J., {Pieri}, M.~M., {Prada}, F., {Price-Whelan}, A.~M., {Raddick}, M.~J.,
  {Rebolo}, R., {Rich}, J., {Richards}, G.~T., {Rockosi}, C.~M., {Roe}, N.~A.,
  {Ross}, A.~J., {Ross}, N.~P., {Rossi}, G., {Rubi{\~n}o-Martin}, J.~A.,
  {Samushia}, L., {S{\'a}nchez}, A.~G., {Sayres}, C., {Schmidt}, S.~J.,
  {Schneider}, D.~P., {Sc{\'o}ccola}, C.~G., {Seo}, H.-J., {Shelden}, A.,
  {Sheldon}, E., {Shen}, Y., {Shu}, Y., {Slosar}, A., {Smee}, S.~A., {Snedden},
  S.~A., {Stauffer}, F., {Steele}, O., {Strauss}, M.~A., {Streblyanska}, A.,
  {Suzuki}, N., {Swanson}, M. E.~C., {Tal}, T., {Tanaka}, M., {Thomas}, D.,
  {Tinker}, J.~L., {Tojeiro}, R., {Tremonti}, C.~A., {Vargas Maga{\~n}a}, M.,
  {Verde}, L., {Viel}, M., {Wake}, D.~A., {Watson}, M., {Weaver}, B.~A.,
  {Weinberg}, D.~H., {Weiner}, B.~J., {West}, A.~A., {White}, M., {Wood-Vasey},
  W.~M., {Yeche}, C., {Zehavi}, I., {Zhao}, G.-B., \& {Zheng}, Z. 2013, \aj,
  145, 10

\bibitem[{{Decarli} {et~al.}(2018){Decarli}, {Walter}, {Venemans},
  {Ba{\~n}ados}, {Bertoldi}, {Carilli}, {Fan}, {Farina}, {Mazzucchelli},
  {Riechers}, {Rix}, {Strauss}, {Wang}, \& {Yang}}]{decarli2018}
{Decarli}, R., {Walter}, F., {Venemans}, B.~P., {Ba{\~n}ados}, E., {Bertoldi},
  F., {Carilli}, C., {Fan}, X., {Farina}, E.~P., {Mazzucchelli}, C.,
  {Riechers}, D., {Rix}, H.-W., {Strauss}, M.~A., {Wang}, R., \& {Yang}, Y.
  2018, \apj, 854, 97

\bibitem[{{Di Matteo} {et~al.}(2017){Di Matteo}, {Croft}, {Feng}, {Waters}, \&
  {Wilkins}}]{matteo2017}
{Di Matteo}, T., {Croft}, R. A.~C., {Feng}, Y., {Waters}, D., \& {Wilkins}, S.
  2017, \mnras, 467, 4243

\bibitem[{{Diemer}(2018)}]{diemer2018}
{Diemer}, B. 2018, \apjs, 239, 35

\bibitem[{{Eilers} {et~al.}(2017){Eilers}, {Davies}, {Hennawi}, {Prochaska},
  {Luki{\'c}}, \& {Mazzucchelli}}]{eilers2017}
{Eilers}, A.-C., {Davies}, F.~B., {Hennawi}, J.~F., {Prochaska}, J.~X.,
  {Luki{\'c}}, Z., \& {Mazzucchelli}, C. 2017, \apj, 840, 24

\bibitem[{{Garc{\'\i}a-Vergara} {et~al.}(2021){Garc{\'\i}a-Vergara}, {Rybak},
  {Hodge}, {Hennawi}, {Decarli}, {Gonz{\'a}lez-L{\'o}pez}, {Arrigoni-Battaia},
  {Aravena}, \& {Farina}}]{garciavergara2021}
{Garc{\'\i}a-Vergara}, C., {Rybak}, M., {Hodge}, J., {Hennawi}, J.~F.,
  {Decarli}, R., {Gonz{\'a}lez-L{\'o}pez}, J., {Arrigoni-Battaia}, F.,
  {Aravena}, M., \& {Farina}, E.~P. 2021, arXiv e-prints, arXiv:2109.09754

\bibitem[{{Gnedin}(2014)}]{gnedin2014}
{Gnedin}, N.~Y. 2014, \apj, 793, 29

\bibitem[{{Gnedin} \& {Abel}(2001)}]{gnedin01}
{Gnedin}, N.~Y. \& {Abel}, T. 2001, \na, 6, 437

\bibitem[{{Habouzit} {et~al.}(2019){Habouzit}, {Volonteri}, {Somerville},
  {Dubois}, {Peirani}, {Pichon}, \& {Devriendt}}]{habouzit2019}
{Habouzit}, M., {Volonteri}, M., {Somerville}, R.~S., {Dubois}, Y., {Peirani},
  S., {Pichon}, C., \& {Devriendt}, J. 2019, \mnras, 489, 1206

\bibitem[{{Inayoshi} {et~al.}(2020){Inayoshi}, {Visbal}, \&
  {Haiman}}]{inayoshi2020}
{Inayoshi}, K., {Visbal}, E., \& {Haiman}, Z. 2020, \araa, 58, 27

\bibitem[{{Kashikawa} {et~al.}(2007){Kashikawa}, {Kitayama}, {Doi}, {Misawa},
  {Komiyama}, \& {Ota}}]{kashikawa2007}
{Kashikawa}, N., {Kitayama}, T., {Doi}, M., {Misawa}, T., {Komiyama}, Y., \&
  {Ota}, K. 2007, \apj, 663, 765

\bibitem[{{Kim} {et~al.}(2009){Kim}, {Stiavelli}, {Trenti}, {Pavlovsky},
  {Djorgovski}, {Scarlata}, {Stern}, {Mahabal}, {Thompson}, {Dickinson},
  {Panagia}, \& {Meylan}}]{kim2009}
{Kim}, S., {Stiavelli}, M., {Trenti}, M., {Pavlovsky}, C.~M., {Djorgovski},
  S.~G., {Scarlata}, C., {Stern}, D., {Mahabal}, A., {Thompson}, D.,
  {Dickinson}, M., {Panagia}, N., \& {Meylan}, G. 2009, \apj, 695, 809

\bibitem[{{Kitayama} {et~al.}(2000){Kitayama}, {Tajiri}, {Umemura}, {Susa}, \&
  {Ikeuchi}}]{kitayama2000}
{Kitayama}, T., {Tajiri}, Y., {Umemura}, M., {Susa}, H., \& {Ikeuchi}, S. 2000,
  \mnras, 315, L1

\bibitem[{{Kravtsov}(1999)}]{kravtsov1999}
{Kravtsov}, A.~V. 1999, PhD thesis, NEW MEXICO STATE UNIVERSITY

\bibitem[{{Kravtsov} {et~al.}(2002){Kravtsov}, {Klypin}, \&
  {Hoffman}}]{kravtsov2002}
{Kravtsov}, A.~V., {Klypin}, A., \& {Hoffman}, Y. 2002, \apj, 571, 563

\bibitem[{{Latif} {et~al.}(2013){Latif}, {Schleicher}, {Schmidt}, \&
  {Niemeyer}}]{latif2013}
{Latif}, M.~A., {Schleicher}, D.~R.~G., {Schmidt}, W., \& {Niemeyer}, J. 2013,
  \mnras, 433, 1607

\bibitem[{{Lupi} {et~al.}(2021){Lupi}, {Haiman}, \& {Volonteri}}]{lupi2021}
{Lupi}, A., {Haiman}, Z., \& {Volonteri}, M. 2021, \mnras, 503, 5046

\bibitem[{{Lusso} {et~al.}(2015){Lusso}, {Worseck}, {Hennawi}, {Prochaska},
  {Vignali}, {Stern}, \& {O'Meara}}]{lusso2015}
{Lusso}, E., {Worseck}, G., {Hennawi}, J.~F., {Prochaska}, J.~X., {Vignali},
  C., {Stern}, J., \& {O'Meara}, J.~M. 2015, \mnras, 449, 4204

\bibitem[{{McGreer} {et~al.}(2014){McGreer}, {Fan}, {Strauss}, {Haiman},
  {Richards}, {Jiang}, {Bian}, \& {Schneider}}]{mcgreer2014}
{McGreer}, I.~D., {Fan}, X., {Strauss}, M.~A., {Haiman}, Z., {Richards}, G.~T.,
  {Jiang}, L., {Bian}, F., \& {Schneider}, D.~P. 2014, \aj, 148, 73

\bibitem[{{Morey} {et~al.}(2021){Morey}, {Eilers}, {Davies}, {Hennawi}, \&
  {Simcoe}}]{morey2021}
{Morey}, K.~A., {Eilers}, A.-C., {Davies}, F.~B., {Hennawi}, J.~F., \&
  {Simcoe}, R.~A. 2021, arXiv e-prints, arXiv:2108.10907

\bibitem[{{Ota} {et~al.}(2018){Ota}, {Venemans}, {Taniguchi}, {Kashikawa},
  {Nakata}, {Harikane}, {Ba{\~n}ados}, {Overzier}, {Riechers}, {Walter},
  {Toshikawa}, {Shibuya}, \& {Jiang}}]{ota2018}
{Ota}, K., {Venemans}, B.~P., {Taniguchi}, Y., {Kashikawa}, N., {Nakata}, F.,
  {Harikane}, Y., {Ba{\~n}ados}, E., {Overzier}, R., {Riechers}, D.~A.,
  {Walter}, F., {Toshikawa}, J., {Shibuya}, T., \& {Jiang}, L. 2018, \apj, 856,
  109

\bibitem[{{P{\^a}ris} {et~al.}(2018){P{\^a}ris}, {Petitjean}, {Aubourg},
  {Myers}, {Streblyanska}, {Lyke}, {Anderson}, {Armengaud}, {Bautista},
  {Blanton}, {Blomqvist}, {Brinkmann}, {Brownstein}, {Brand t}, {Burtin},
  {Dawson}, {de la Torre}, {Georgakakis}, {Gil-Mar{\'\i}n}, {Green}, {Hall},
  {Kneib}, {LaMassa}, {Le Goff}, {MacLeod}, {Mariappan}, {McGreer}, {Merloni},
  {Noterdaeme}, {Palanque-Delabrouille}, {Percival}, {Ross}, {Rossi},
  {Schneider}, {Seo}, {Tojeiro}, {Weaver}, {Weijmans}, {Y{\`e}che}, {Zarrouk},
  \& {Zhao}}]{paris2018}
{P{\^a}ris}, I., {Petitjean}, P., {Aubourg}, {\'E}., {Myers}, A.~D.,
  {Streblyanska}, A., {Lyke}, B.~W., {Anderson}, S.~F., {Armengaud}, {\'E}.,
  {Bautista}, J., {Blanton}, M.~R., {Blomqvist}, M., {Brinkmann}, J.,
  {Brownstein}, J.~R., {Brand t}, W.~N., {Burtin}, {\'E}., {Dawson}, K., {de la
  Torre}, S., {Georgakakis}, A., {Gil-Mar{\'\i}n}, H., {Green}, P.~J., {Hall},
  P.~B., {Kneib}, J.-P., {LaMassa}, S.~M., {Le Goff}, J.-M., {MacLeod}, C.,
  {Mariappan}, V., {McGreer}, I.~D., {Merloni}, A., {Noterdaeme}, P.,
  {Palanque-Delabrouille}, N., {Percival}, W.~J., {Ross}, A.~J., {Rossi}, G.,
  {Schneider}, D.~P., {Seo}, H.-J., {Tojeiro}, R., {Weaver}, B.~A., {Weijmans},
  A.-M., {Y{\`e}che}, C., {Zarrouk}, P., \& {Zhao}, G.-B. 2018, \aap, 613, A51

\bibitem[{{P{\^a}ris} {et~al.}(2011){P{\^a}ris}, {Petitjean}, {Rollinde},
  {Aubourg}, {Busca}, {Charlassier}, {Delubac}, {Hamilton}, {Le Goff},
  {Palanque-Delabrouille}, {Peirani}, {Pichon}, {Rich}, {Vargas-Maga{\~n}a}, \&
  {Y{\`e}che}}]{paris2011}
{P{\^a}ris}, I., {Petitjean}, P., {Rollinde}, E., {Aubourg}, E., {Busca}, N.,
  {Charlassier}, R., {Delubac}, T., {Hamilton}, J.-C., {Le Goff}, J.-M.,
  {Palanque-Delabrouille}, N., {Peirani}, S., {Pichon}, C., {Rich}, J.,
  {Vargas-Maga{\~n}a}, M., \& {Y{\`e}che}, C. 2011, \aap, 530, A50

\bibitem[{{Planck Collaboration} {et~al.}(2020){Planck Collaboration},
  {Aghanim}, {Akrami}, {Ashdown}, {Aumont}, {Baccigalupi}, {Ballardini},
  {Banday}, {Barreiro}, {Bartolo}, {Basak}, {Battye}, {Benabed}, {Bernard},
  {Bersanelli}, {Bielewicz}, {Bock}, {Bond}, {Borrill}, {Bouchet}, {Boulanger},
  {Bucher}, {Burigana}, {Butler}, {Calabrese}, {Cardoso}, {Carron},
  {Challinor}, {Chiang}, {Chluba}, {Colombo}, {Combet}, {Contreras}, {Crill},
  {Cuttaia}, {de Bernardis}, {de Zotti}, {Delabrouille}, {Delouis}, {Di
  Valentino}, {Diego}, {Dor{\'e}}, {Douspis}, {Ducout}, {Dupac}, {Dusini},
  {Efstathiou}, {Elsner}, {En{\ss}lin}, {Eriksen}, {Fantaye}, {Farhang},
  {Fergusson}, {Fernandez-Cobos}, {Finelli}, {Forastieri}, {Frailis},
  {Fraisse}, {Franceschi}, {Frolov}, {Galeotta}, {Galli}, {Ganga},
  {G{\'e}nova-Santos}, {Gerbino}, {Ghosh}, {Gonz{\'a}lez-Nuevo}, {G{\'o}rski},
  {Gratton}, {Gruppuso}, {Gudmundsson}, {Hamann}, {Handley}, {Hansen},
  {Herranz}, {Hildebrandt}, {Hivon}, {Huang}, {Jaffe}, {Jones}, {Karakci},
  {Keih{\"a}nen}, {Keskitalo}, {Kiiveri}, {Kim}, {Kisner}, {Knox},
  {Krachmalnicoff}, {Kunz}, {Kurki-Suonio}, {Lagache}, {Lamarre}, {Lasenby},
  {Lattanzi}, {Lawrence}, {Le Jeune}, {Lemos}, {Lesgourgues}, {Levrier},
  {Lewis}, {Liguori}, {Lilje}, {Lilley}, {Lindholm}, {L{\'o}pez-Caniego},
  {Lubin}, {Ma}, {Mac{\'\i}as-P{\'e}rez}, {Maggio}, {Maino}, {Mandolesi},
  {Mangilli}, {Marcos-Caballero}, {Maris}, {Martin}, {Martinelli},
  {Mart{\'\i}nez-Gonz{\'a}lez}, {Matarrese}, {Mauri}, {McEwen}, {Meinhold},
  {Melchiorri}, {Mennella}, {Migliaccio}, {Millea}, {Mitra},
  {Miville-Desch{\^e}nes}, {Molinari}, {Montier}, {Morgante}, {Moss}, {Natoli},
  {N{\o}rgaard-Nielsen}, {Pagano}, {Paoletti}, {Partridge}, {Patanchon},
  {Peiris}, {Perrotta}, {Pettorino}, {Piacentini}, {Polastri}, {Polenta},
  {Puget}, {Rachen}, {Reinecke}, {Remazeilles}, {Renzi}, {Rocha}, {Rosset},
  {Roudier}, {Rubi{\~n}o-Mart{\'\i}n}, {Ruiz-Granados}, {Salvati}, {Sandri},
  {Savelainen}, {Scott}, {Shellard}, {Sirignano}, {Sirri}, {Spencer},
  {Sunyaev}, {Suur-Uski}, {Tauber}, {Tavagnacco}, {Tenti}, {Toffolatti},
  {Tomasi}, {Trombetti}, {Valenziano}, {Valiviita}, {Van Tent}, {Vibert},
  {Vielva}, {Villa}, {Vittorio}, {Wandelt}, {Wehus}, {White}, {White},
  {Zacchei}, \& {Zonca}}]{planck2020}
{Planck Collaboration}, {Aghanim}, N., {Akrami}, Y., {Ashdown}, M., {Aumont},
  J., {Baccigalupi}, C., {Ballardini}, M., {Banday}, A.~J., {Barreiro}, R.~B.,
  {Bartolo}, N., {Basak}, S., {Battye}, R., {Benabed}, K., {Bernard}, J.~P.,
  {Bersanelli}, M., {Bielewicz}, P., {Bock}, J.~J., {Bond}, J.~R., {Borrill},
  J., {Bouchet}, F.~R., {Boulanger}, F., {Bucher}, M., {Burigana}, C.,
  {Butler}, R.~C., {Calabrese}, E., {Cardoso}, J.~F., {Carron}, J.,
  {Challinor}, A., {Chiang}, H.~C., {Chluba}, J., {Colombo}, L.~P.~L.,
  {Combet}, C., {Contreras}, D., {Crill}, B.~P., {Cuttaia}, F., {de Bernardis},
  P., {de Zotti}, G., {Delabrouille}, J., {Delouis}, J.~M., {Di Valentino}, E.,
  {Diego}, J.~M., {Dor{\'e}}, O., {Douspis}, M., {Ducout}, A., {Dupac}, X.,
  {Dusini}, S., {Efstathiou}, G., {Elsner}, F., {En{\ss}lin}, T.~A., {Eriksen},
  H.~K., {Fantaye}, Y., {Farhang}, M., {Fergusson}, J., {Fernandez-Cobos}, R.,
  {Finelli}, F., {Forastieri}, F., {Frailis}, M., {Fraisse}, A.~A.,
  {Franceschi}, E., {Frolov}, A., {Galeotta}, S., {Galli}, S., {Ganga}, K.,
  {G{\'e}nova-Santos}, R.~T., {Gerbino}, M., {Ghosh}, T., {Gonz{\'a}lez-Nuevo},
  J., {G{\'o}rski}, K.~M., {Gratton}, S., {Gruppuso}, A., {Gudmundsson}, J.~E.,
  {Hamann}, J., {Handley}, W., {Hansen}, F.~K., {Herranz}, D., {Hildebrandt},
  S.~R., {Hivon}, E., {Huang}, Z., {Jaffe}, A.~H., {Jones}, W.~C., {Karakci},
  A., {Keih{\"a}nen}, E., {Keskitalo}, R., {Kiiveri}, K., {Kim}, J., {Kisner},
  T.~S., {Knox}, L., {Krachmalnicoff}, N., {Kunz}, M., {Kurki-Suonio}, H.,
  {Lagache}, G., {Lamarre}, J.~M., {Lasenby}, A., {Lattanzi}, M., {Lawrence},
  C.~R., {Le Jeune}, M., {Lemos}, P., {Lesgourgues}, J., {Levrier}, F.,
  {Lewis}, A., {Liguori}, M., {Lilje}, P.~B., {Lilley}, M., {Lindholm}, V.,
  {L{\'o}pez-Caniego}, M., {Lubin}, P.~M., {Ma}, Y.~Z.,
  {Mac{\'\i}as-P{\'e}rez}, J.~F., {Maggio}, G., {Maino}, D., {Mandolesi}, N.,
  {Mangilli}, A., {Marcos-Caballero}, A., {Maris}, M., {Martin}, P.~G.,
  {Martinelli}, M., {Mart{\'\i}nez-Gonz{\'a}lez}, E., {Matarrese}, S., {Mauri},
  N., {McEwen}, J.~D., {Meinhold}, P.~R., {Melchiorri}, A., {Mennella}, A.,
  {Migliaccio}, M., {Millea}, M., {Mitra}, S., {Miville-Desch{\^e}nes}, M.~A.,
  {Molinari}, D., {Montier}, L., {Morgante}, G., {Moss}, A., {Natoli}, P.,
  {N{\o}rgaard-Nielsen}, H.~U., {Pagano}, L., {Paoletti}, D., {Partridge}, B.,
  {Patanchon}, G., {Peiris}, H.~V., {Perrotta}, F., {Pettorino}, V.,
  {Piacentini}, F., {Polastri}, L., {Polenta}, G., {Puget}, J.~L., {Rachen},
  J.~P., {Reinecke}, M., {Remazeilles}, M., {Renzi}, A., {Rocha}, G., {Rosset},
  C., {Roudier}, G., {Rubi{\~n}o-Mart{\'\i}n}, J.~A., {Ruiz-Granados}, B.,
  {Salvati}, L., {Sandri}, M., {Savelainen}, M., {Scott}, D., {Shellard},
  E.~P.~S., {Sirignano}, C., {Sirri}, G., {Spencer}, L.~D., {Sunyaev}, R.,
  {Suur-Uski}, A.~S., {Tauber}, J.~A., {Tavagnacco}, D., {Tenti}, M.,
  {Toffolatti}, L., {Tomasi}, M., {Trombetti}, T., {Valenziano}, L.,
  {Valiviita}, J., {Van Tent}, B., {Vibert}, L., {Vielva}, P., {Villa}, F.,
  {Vittorio}, N., {Wandelt}, B.~D., {Wehus}, I.~K., {White}, M., {White},
  S.~D.~M., {Zacchei}, A., \& {Zonca}, A. 2020, \aap, 641, A6

\bibitem[{{Porredon} {et~al.}(2021){Porredon}, {Crocce}, {Fosalba},
  {Elvin-Poole}, {Carnero Rosell}, {Cawthon}, {Eifler}, {Fang}, {Ferrero},
  {Krause}, {MacCrann}, {Weaverdyck}, {Abbott}, {Aguena}, {Allam}, {Amon},
  {Avila}, {Bacon}, {Bertin}, {Bhargava}, {Bridle}, {Brooks}, {Carrasco Kind},
  {Carretero}, {Castander}, {Choi}, {Costanzi}, {da Costa}, {Pereira}, {De
  Vicente}, {Desai}, {Diehl}, {Doel}, {Drlica-Wagner}, {Eckert}, {Fert{\'e}},
  {Flaugher}, {Frieman}, {Garc{\'\i}a-Bellido}, {Gaztanaga}, {Gerdes},
  {Giannantonio}, {Gruen}, {Gruendl}, {Gschwend}, {Gutierrez}, {Hartley},
  {Hinton}, {Hollowood}, {Honscheid}, {Hoyle}, {James}, {Jarvis}, {Kuehn},
  {Kuropatkin}, {Maia}, {Marshall}, {Menanteau}, {Miquel}, {Morgan}, {Palmese},
  {Pandey}, {Paz-Chinch{\'o}n}, {Plazas}, {Rodriguez-Monroy}, {Roodman},
  {Samuroff}, {Sanchez}, {Scarpine}, {Serrano}, {Sevilla-Noarbe}, {Smith},
  {Soares-Santos}, {Suchyta}, {Swanson}, {Tarle}, {To}, {Varga}, {Weller},
  {Wilkinson}, \& {DES Collaboration}}]{des2021}
{Porredon}, A., {Crocce}, M., {Fosalba}, P., {Elvin-Poole}, J., {Carnero
  Rosell}, A., {Cawthon}, R., {Eifler}, T.~F., {Fang}, X., {Ferrero}, I.,
  {Krause}, E., {MacCrann}, N., {Weaverdyck}, N., {Abbott}, T.~M.~C., {Aguena},
  M., {Allam}, S., {Amon}, A., {Avila}, S., {Bacon}, D., {Bertin}, E.,
  {Bhargava}, S., {Bridle}, S.~L., {Brooks}, D., {Carrasco Kind}, M.,
  {Carretero}, J., {Castander}, F.~J., {Choi}, A., {Costanzi}, M., {da Costa},
  L.~N., {Pereira}, M.~E.~S., {De Vicente}, J., {Desai}, S., {Diehl}, H.~T.,
  {Doel}, P., {Drlica-Wagner}, A., {Eckert}, K., {Fert{\'e}}, A., {Flaugher},
  B., {Frieman}, J., {Garc{\'\i}a-Bellido}, J., {Gaztanaga}, E., {Gerdes},
  D.~W., {Giannantonio}, T., {Gruen}, D., {Gruendl}, R.~A., {Gschwend}, J.,
  {Gutierrez}, G., {Hartley}, W.~G., {Hinton}, S.~R., {Hollowood}, D.~L.,
  {Honscheid}, K., {Hoyle}, B., {James}, D.~J., {Jarvis}, M., {Kuehn}, K.,
  {Kuropatkin}, N., {Maia}, M.~A.~G., {Marshall}, J.~L., {Menanteau}, F.,
  {Miquel}, R., {Morgan}, R., {Palmese}, A., {Pandey}, S., {Paz-Chinch{\'o}n},
  F., {Plazas}, A.~A., {Rodriguez-Monroy}, M., {Roodman}, A., {Samuroff}, S.,
  {Sanchez}, E., {Scarpine}, V., {Serrano}, S., {Sevilla-Noarbe}, I., {Smith},
  M., {Soares-Santos}, M., {Suchyta}, E., {Swanson}, M.~E.~C., {Tarle}, G.,
  {To}, C., {Varga}, T.~N., {Weller}, J., {Wilkinson}, R.~D., \& {DES
  Collaboration}. 2021, \prd, 103, 043503

\bibitem[{{Regan} \& {Haehnelt}(2009)}]{regan2009}
{Regan}, J.~A. \& {Haehnelt}, M.~G. 2009, \mnras, 396, 343

\bibitem[{{Rudd} {et~al.}(2008){Rudd}, {Zentner}, \& {Kravtsov}}]{rudd2008}
{Rudd}, D.~H., {Zentner}, A.~R., \& {Kravtsov}, A.~V. 2008, \apj, 672, 19

\bibitem[{{Runnoe} {et~al.}(2012){Runnoe}, {Brotherton}, \&
  {Shang}}]{runnoe2012}
{Runnoe}, J.~C., {Brotherton}, M.~S., \& {Shang}, Z. 2012, \mnras, 422, 478

\bibitem[{{Schindler} {et~al.}(2020){Schindler}, {Farina}, {Ba{\~n}ados},
  {Eilers}, {Hennawi}, {Onoue}, {Venemans}, {Walter}, {Wang}, {Davies},
  {Decarli}, {Rosa}, {Drake}, {Fan}, {Mazzucchelli}, {Rix}, {Worseck}, \&
  {Yang}}]{schindler2020}
{Schindler}, J.-T., {Farina}, E.~P., {Ba{\~n}ados}, E., {Eilers}, A.-C.,
  {Hennawi}, J.~F., {Onoue}, M., {Venemans}, B.~P., {Walter}, F., {Wang}, F.,
  {Davies}, F.~B., {Decarli}, R., {Rosa}, G.~D., {Drake}, A., {Fan}, X.,
  {Mazzucchelli}, C., {Rix}, H.-W., {Worseck}, G., \& {Yang}, J. 2020, \apj,
  905, 51

\bibitem[{{Shang} {et~al.}(2010){Shang}, {Bryan}, \& {Haiman}}]{shang2010}
{Shang}, C., {Bryan}, G.~L., \& {Haiman}, Z. 2010, \mnras, 402, 1249

\bibitem[{{Simpson} {et~al.}(2014){Simpson}, {Mortlock}, {Warren}, {Cantalupo},
  {Hewett}, {McLure}, {McMahon}, \& {Venemans}}]{simpson2014}
{Simpson}, C., {Mortlock}, D., {Warren}, S., {Cantalupo}, S., {Hewett}, P.,
  {McLure}, R., {McMahon}, R., \& {Venemans}, B. 2014, \mnras, 442, 3454

\bibitem[{{Venemans} {et~al.}(2020){Venemans}, {Walter}, {Neeleman}, {Novak},
  {Otter}, {Decarli}, {Ba{\~n}ados}, {Drake}, {Farina}, {Kaasinen},
  {Mazzucchelli}, {Carilli}, {Fan}, {Rix}, \& {Wang}}]{venemans2020}
{Venemans}, B.~P., {Walter}, F., {Neeleman}, M., {Novak}, M., {Otter}, J.,
  {Decarli}, R., {Ba{\~n}ados}, E., {Drake}, A., {Farina}, E.~P., {Kaasinen},
  M., {Mazzucchelli}, C., {Carilli}, C., {Fan}, X., {Rix}, H.-W., \& {Wang}, R.
  2020, \apj, 904, 130

\bibitem[{{Vernet} {et~al.}(2011){Vernet}, {Dekker}, {D'Odorico}, {Kaper},
  {Kjaergaard}, {Hammer}, {Randich}, {Zerbi}, {Groot}, {Hjorth}, {Guinouard},
  {Navarro}, {Adolfse}, {Albers}, {Amans}, {Andersen}, {Andersen}, {Binetruy},
  {Bristow}, {Castillo}, {Chemla}, {Christensen}, {Conconi}, {Conzelmann},
  {Dam}, {de Caprio}, {de Ugarte Postigo}, {Delabre}, {di Marcantonio},
  {Downing}, {Elswijk}, {Finger}, {Fischer}, {Flores}, {Fran{\c{c}}ois},
  {Goldoni}, {Guglielmi}, {Haigron}, {Hanenburg}, {Hendriks}, {Horrobin},
  {Horville}, {Jessen}, {Kerber}, {Kern}, {Kiekebusch}, {Kleszcz}, {Klougart},
  {Kragt}, {Larsen}, {Lizon}, {Lucuix}, {Mainieri}, {Manuputy}, {Martayan},
  {Mason}, {Mazzoleni}, {Michaelsen}, {Modigliani}, {Moehler}, {M{\o}ller},
  {Norup S{\o}rensen}, {N{\o}rregaard}, {P{\'e}roux}, {Patat}, {Pena}, {Pragt},
  {Reinero}, {Rigal}, {Riva}, {Roelfsema}, {Royer}, {Sacco}, {Santin},
  {Schoenmaker}, {Spano}, {Sweers}, {Ter Horst}, {Tintori}, {Tromp}, {van
  Dael}, {van der Vliet}, {Venema}, {Vidali}, {Vinther}, {Vola}, {Winters},
  {Wistisen}, {Wulterkens}, \& {Zacchei}}]{vernet2011}
{Vernet}, J., {Dekker}, H., {D'Odorico}, S., {Kaper}, L., {Kjaergaard}, P.,
  {Hammer}, F., {Randich}, S., {Zerbi}, F., {Groot}, P.~J., {Hjorth}, J.,
  {Guinouard}, I., {Navarro}, R., {Adolfse}, T., {Albers}, P.~W., {Amans},
  J.~P., {Andersen}, J.~J., {Andersen}, M.~I., {Binetruy}, P., {Bristow}, P.,
  {Castillo}, R., {Chemla}, F., {Christensen}, L., {Conconi}, P., {Conzelmann},
  R., {Dam}, J., {de Caprio}, V., {de Ugarte Postigo}, A., {Delabre}, B., {di
  Marcantonio}, P., {Downing}, M., {Elswijk}, E., {Finger}, G., {Fischer}, G.,
  {Flores}, H., {Fran{\c{c}}ois}, P., {Goldoni}, P., {Guglielmi}, L.,
  {Haigron}, R., {Hanenburg}, H., {Hendriks}, I., {Horrobin}, M., {Horville},
  D., {Jessen}, N.~C., {Kerber}, F., {Kern}, L., {Kiekebusch}, M., {Kleszcz},
  P., {Klougart}, J., {Kragt}, J., {Larsen}, H.~H., {Lizon}, J.~L., {Lucuix},
  C., {Mainieri}, V., {Manuputy}, R., {Martayan}, C., {Mason}, E., {Mazzoleni},
  R., {Michaelsen}, N., {Modigliani}, A., {Moehler}, S., {M{\o}ller}, P.,
  {Norup S{\o}rensen}, A., {N{\o}rregaard}, P., {P{\'e}roux}, C., {Patat}, F.,
  {Pena}, E., {Pragt}, J., {Reinero}, C., {Rigal}, F., {Riva}, M., {Roelfsema},
  R., {Royer}, F., {Sacco}, G., {Santin}, P., {Schoenmaker}, T., {Spano}, P.,
  {Sweers}, E., {Ter Horst}, R., {Tintori}, M., {Tromp}, N., {van Dael}, P.,
  {van der Vliet}, H., {Venema}, L., {Vidali}, M., {Vinther}, J., {Vola}, P.,
  {Winters}, R., {Wistisen}, D., {Wulterkens}, G., \& {Zacchei}, A. 2011, \aap,
  536, A105

\bibitem[{{Wang} {et~al.}(2021{\natexlab{a}}){Wang}, {Fan}, {Yang},
  {Mazzucchelli}, {Wu}, {Li}, {Ba{\~n}ados}, {Farina}, {Nanni}, {Ai}, {Bian},
  {Davies}, {Decarli}, {Hennawi}, {Schindler}, {Venemans}, \&
  {Walter}}]{wang2021b}
{Wang}, F., {Fan}, X., {Yang}, J., {Mazzucchelli}, C., {Wu}, X.-B., {Li},
  J.-T., {Ba{\~n}ados}, E., {Farina}, E.~P., {Nanni}, R., {Ai}, Y., {Bian}, F.,
  {Davies}, F.~B., {Decarli}, R., {Hennawi}, J.~F., {Schindler}, J.-T.,
  {Venemans}, B., \& {Walter}, F. 2021{\natexlab{a}}, \apj, 908, 53

\bibitem[{{Wang} {et~al.}(2021{\natexlab{b}}){Wang}, {Yang}, {Fan}, {Hennawi},
  {Barth}, {Banados}, {Bian}, {Boutsia}, {Connor}, {Davies}, {Decarli},
  {Eilers}, {Farina}, {Green}, {Jiang}, {Li}, {Mazzucchelli}, {Nanni},
  {Schindler}, {Venemans}, {Walter}, {Wu}, \& {Yue}}]{wang2021}
{Wang}, F., {Yang}, J., {Fan}, X., {Hennawi}, J.~F., {Barth}, A.~J., {Banados},
  E., {Bian}, F., {Boutsia}, K., {Connor}, T., {Davies}, F.~B., {Decarli}, R.,
  {Eilers}, A.-C., {Farina}, E.~P., {Green}, R., {Jiang}, L., {Li}, J.-T.,
  {Mazzucchelli}, C., {Nanni}, R., {Schindler}, J.-T., {Venemans}, B.,
  {Walter}, F., {Wu}, X.-B., \& {Yue}, M. 2021{\natexlab{b}}, \apjl, 907, L1

\bibitem[{{Wang} {et~al.}(2010){Wang}, {Carilli}, {Neri}, {Riechers}, {Wagg},
  {Walter}, {Bertoldi}, {Menten}, {Omont}, {Cox}, \& {Fan}}]{wang2010}
{Wang}, R., {Carilli}, C.~L., {Neri}, R., {Riechers}, D.~A., {Wagg}, J.,
  {Walter}, F., {Bertoldi}, F., {Menten}, K.~M., {Omont}, A., {Cox}, P., \&
  {Fan}, X. 2010, \apj, 714, 699

\bibitem[{{Wise} {et~al.}(2008){Wise}, {Turk}, \& {Abel}}]{wise2008}
{Wise}, J.~H., {Turk}, M.~J., \& {Abel}, T. 2008, \apj, 682, 745

\bibitem[{{Wu} {et~al.}(2020){Wu}, {Motloch}, {Hu}, \& {Raveri}}]{wu2020}
{Wu}, W.~L.~K., {Motloch}, P., {Hu}, W., \& {Raveri}, M. 2020, \prd, 102,
  023510

\bibitem[{{Yang} {et~al.}(2020){Yang}, {Wang}, {Fan}, {Hennawi}, {Davies},
  {Yue}, {Banados}, {Wu}, {Venemans}, {Barth}, {Bian}, {Boutsia}, {Decarli},
  {Farina}, {Green}, {Jiang}, {Li}, {Mazzucchelli}, \& {Walter}}]{yang2020}
{Yang}, J., {Wang}, F., {Fan}, X., {Hennawi}, J.~F., {Davies}, F.~B., {Yue},
  M., {Banados}, E., {Wu}, X.-B., {Venemans}, B., {Barth}, A.~J., {Bian}, F.,
  {Boutsia}, K., {Decarli}, R., {Farina}, E.~P., {Green}, R., {Jiang}, L.,
  {Li}, J.-T., {Mazzucchelli}, C., \& {Walter}, F. 2020, \apjl, 897, L14

\bibitem[{{Young} {et~al.}(1979){Young}, {Sargent}, {Boksenberg}, {Carswell},
  \& {Whelan}}]{young1979}
{Young}, P.~J., {Sargent}, W.~L.~W., {Boksenberg}, A., {Carswell}, R.~F., \&
  {Whelan}, J.~A.~J. 1979, \apj, 229, 891

\bibitem[{{Zhu} {et~al.}(2021){Zhu}, {Becker}, {Bosman}, {Keating},
  {Christenson}, {Ba{\~n}ados}, {Bian}, {Davies}, {D'Odorico}, {Eilers}, {Fan},
  {Haehnelt}, {Kulkarni}, {Pallottini}, {Qin}, {Wang}, \& {Yang}}]{zhu2021}
{Zhu}, Y., {Becker}, G.~D., {Bosman}, S. E.~I., {Keating}, L.~C.,
  {Christenson}, H.~M., {Ba{\~n}ados}, E., {Bian}, F., {Davies}, F.~B.,
  {D'Odorico}, V., {Eilers}, A.-C., {Fan}, X., {Haehnelt}, M.~G., {Kulkarni},
  G., {Pallottini}, A., {Qin}, Y., {Wang}, F., \& {Yang}, J. 2021, arXiv
  e-prints, arXiv:2109.06295

\end{thebibliography}

\end{document}